\documentstyle[emulateapj,psfig]{article}   
\makeatletter

\newenvironment{inlinefigure}{%
\def\@captype{figure}%
\noindent\begin{minipage}{0.999\linewidth}\begin{center}}
{\end{center}\end{minipage}\smallskip}
\makeatother

\lefthead{HELLAS -V: THE X-RAY POPULATION AND EVOLUTION}
\righthead{LA FRANCA ET AL.}

\newcommand{\ecs}{erg cm$^{-2}$ s$^{-1}$}

\begin{document}
\title{The {\it BeppoSAX} High Energy
Large Area Survey -- V. The nature of the hard X-ray source populations
and its evolution}
\author{F.\ La Franca,$\!$\altaffilmark{1}
F.\ Fiore,$\!$\altaffilmark{2}
C.\ Vignali,$\!$\altaffilmark{3}
A.\ Antonelli,$\!$\altaffilmark{2}
A.\ Comastri,$\!$\altaffilmark{4}
P.\ Giommi,$\!$\altaffilmark{5}
G.\ Matt,$\!$\altaffilmark{1}
S.\ Molendi,$\!$\altaffilmark{6}
G.\,C.\ Perola,$\!$\altaffilmark{1}
F.\ Pompilio\ $\!$\altaffilmark{7}
}
\altaffiltext{1}{Dipartimento di Fisica, Universit\`a degli Studi Roma Tre,
Via della Vasca Navale 84, I-00146 Roma, Italy }
\altaffiltext{2}{Osservatorio Astronomico di Roma, Via Frascati 33, I-00040 Monteporzio, Italy}
\altaffiltext{3}{Dep. of Astronomy and Astrophysics, The Pennsylvania State University, University Park, PA 16802, USA}
\altaffiltext{4}{Osservatorio Astronomico di Bologna, Via Ranzani 1, I-40127 Bologna, Italy}
\altaffiltext{5}{ASI Science Data Center, Via Galileo Galilei, I-00044 Frascati, Italy}
\altaffiltext{6}{Istituto di Fisica Cosmica - CNR, Via Bassini 15, I-20121 Milano, Italy}
\altaffiltext{7}{SISSA, Via Beirut 4, I-34014 Trieste, Italy}


\slugcomment{Received 2001 October 31; Accepted 2002 January 10;
To appear on ApJ, volume 570, May 1, 2002}

\begin{abstract}
              
We present optical spectroscopic identifications of hard X-ray (5-10
keV) selected sources belonging to the HELLAS sample obtained with
{\it BeppoSAX} down to a 5-10 keV flux limit of $f_{5-10
keV}$$\sim$$3\times 10^{-14}$ \ecs. The sample consists
of 118 sources. 25 sources have been identified trough correlations
with catalogues of known sources. 49 have been searched for
spectroscopic identification at the telescope. 13 fields resulted
empty down to R=21. 37 sources have been identified with type 1 AGN
and 9 with type 2 AGN. The remaining are: 5 narrow emission line
galaxies, 6 Clusters, 2 BL Lac, 1 Radio Galaxy and 1 Star. Combining
these objects with other hard X-ray selected AGNs from {\it ASCA} and
{\it HEAO1}, we find that the local luminosity function of type 1 AGN
in the 2-10 keV band is fairly well represented by a
double-power-law-function. There is evidence for significant
cosmological evolution according to a pure luminosity evolution (PLE)
model $L_X(z)$$\propto$$(1+z)^k$, with $k$=2.12 and $k$=2.22
($\sigma_k$$\simeq$$0.14$) in a
($\Omega_m$,$\Omega_\lambda$)=(1.0,0.0) and in a
($\Omega_m$,$\Omega_\lambda$)=(0.3,0.7) cosmology, respectively. The
data show an excess of faint high redshift type 1 AGN which is well
modeled by a luminosity dependent density evolution (LDDE), similarly
to what observed in the soft X-rays. However, in both cosmologies, the
statistic is not significant enough to distinguish between the PLE and
LDDE models. The fitted models imply a contribution of AGN1 to the
2-10 keV X-ray background from 35\% up to 60\%.

\end{abstract}

\keywords{cosmology: observations --- galaxies: distances and
redshift --- galaxies: evolution --- galaxies: active --- quasars:
general}

\section{Introduction}
\label{secintro}

AGN have first been discovered in the radio and soon after searched in
the optical band. Consequently, they have been classified using their
optical characteristics and mainly divided into two categories: type 1
(AGN1) and 2 (AGN2) according to the presence or not of broad emission
lines in their optical spectra (we will keep this definition of AGN1
throughout this paper).

Before the advent of the last generation of hard X-ray telescopes, AGN
samples where predominantly based on AGN1 selected either in the
optical or, later on, in the soft X-rays by {\it Einstein} and {\it
ROSAT}. In these bands the evolution of AGN1 has been well measured
(see e.g.  Della Ceca et al. 1992; Boyle et al. 2000; Miyaji,
Hasinger, \& Schmidt 2000). On the contrary the production of samples
of AGN2 has been difficult at any wavelength and limited to few local
surveys.

The general picture was in favor of a model in which AGN1 objects were
associated to AGN with low absorption in the hard X-rays while AGN2 to
obscured sources with large column densities and spectra strongly
depressed in the soft X-rays, as expected in the unification models
(e.g. Antonucci 1993).

In the last decade the advent of the {\it ASCA} and {\it BeppoSAX}
satellites has allowed for the first time the detection and
identification of AGN as the main counterparts of hard (2-10 keV)
X-ray sources down to fluxes $\sim 5\times 10^{-14}$ \ecs, more than 2
orders of magnitude fainter than {\it HEAO1} (Wood et al. 1984). These
identifications accounted for about 30\% of the 2-10 keV hard X-ray
background (Ueda et al. 1998; Fiore et al. 1999). Recently the new
generation of X-ray satellites such as {\it Chandra} and {\it
XMM-Newton}, have reached fluxes 100 times fainter, identifying
hundreds of sources and almost resolving the hard (2-10 keV) X-ray
background (e.g. Mushotzky et al. 2000; Fiore et al. 2000; Giacconi et
al. 2001; Hornschemeier et al. 2001; Hasinger et al. 2001;
Tozzi et al. 2001; Baldi et al. 2001).

Thanks to their excellent angular resolution ($\sim$1-5$''$), the
first spectroscopic identifications projects have been able to observe
faint (I$\sim$23) optical counterparts. At variance with the classical
type-1/type-2 model in the optical, a significant number of the
counterparts ($\sim$30\%) resulted to be apparently optical normal
galaxies, with X-ray luminosities $L_X$$\simeq$$10^{42}-10^{44}$
erg s$^{-1}$ typical of AGN activity, and moreover part of the
optical type 1 AGNs resulted to be absorbed in the hard X-rays (see
e.g. Fiore et al. 2000; Barger et al. 2001; Tozzi et al. 2001;
Hornschemeier et al. 2001; Comastri et al. 2002).

These observations have complicated the picture of the AGN model. In
this framework the computation of the density of AGN has become an
even more difficult task. In fact, it is not clear how to classify the
sources and to take into account the selection biases introduced by
the observation in the 2-10 keV range, where the absorption still play
a relevant role.

%
%
\begin{inlinefigure}
\psfig{figure=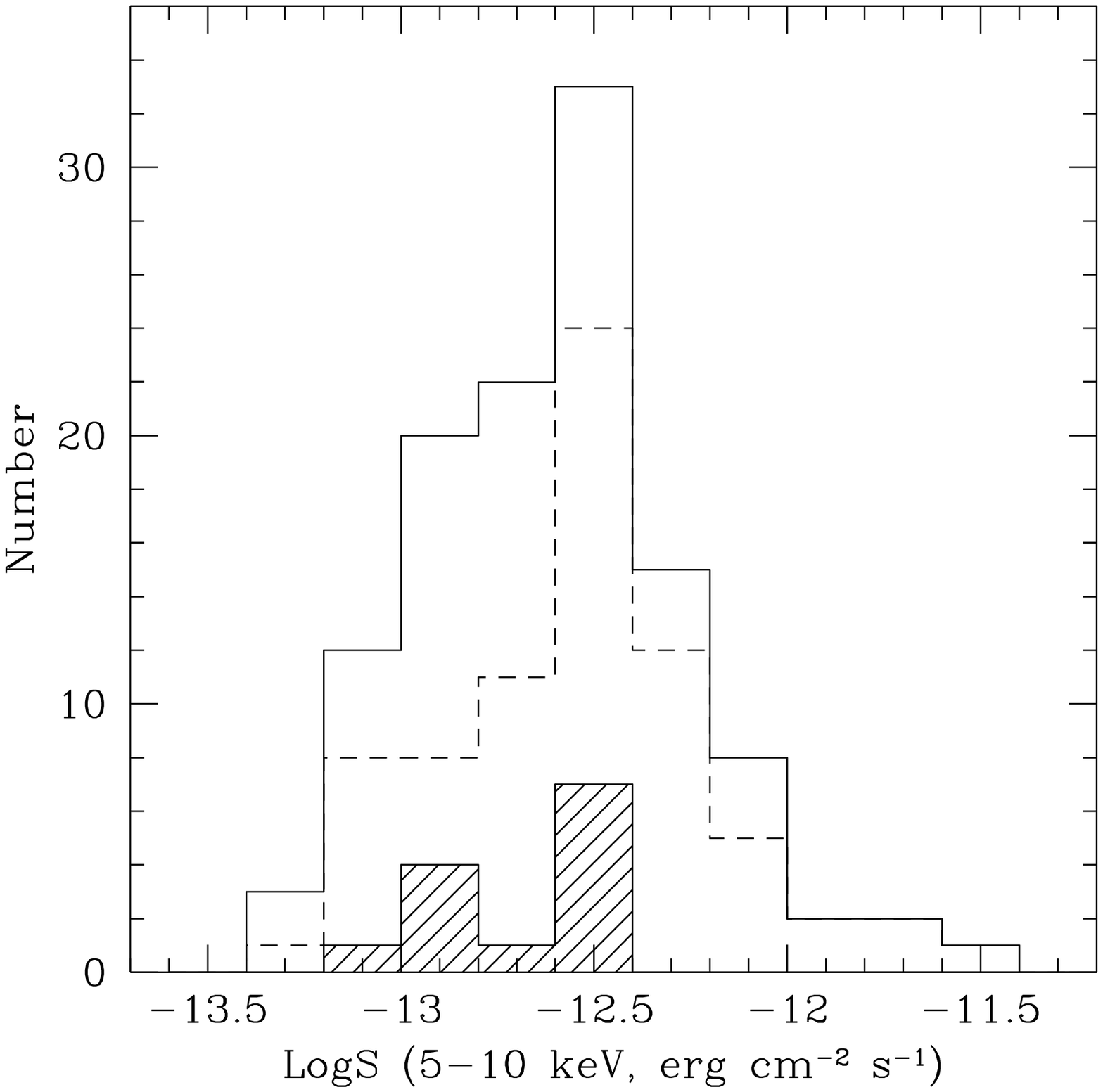,angle=0,width=8.1cm}
\caption{
The flux distribution of the 118 sources from HELLAS used in this
analysis (continuous line). The dashed line is the distribution of the
74 sources for which a spectroscopic identification campaign has been
carried out ({\it empty} fields included).  The distribution of the 13
sources for which we end up with an {\it empty} field is represented by the
hatched histogram.
}
\figurenum{1}
\addtolength{\baselineskip}{5pt}
\end{inlinefigure}

These recent deep surveys with {\it Chandra} and {\it XMM-Newton} have
reached fluxes $\sim$ $5\times 10^{-16}$ \ecs (2-10 keV) in quite
small areas (less than 1 deg$^2$). As a consequence these surveys are
not able to provide statistical significant samples at brighter fluxes
($\sim 10^{-13}$ \ecs; 5-10 keV) where the density of sources is about
5/deg$^2$ (Fiore et al. 2001) and tens of square degrees are to be
covered. Such data are necessary to provide large numbers of
spectroscopic identified sources in a wide range of X-ray fluxes in
order to cover as much as possible the $L_X/z$ plane and hence to
derive their X-ray luminosity function (LF).

In this paper we report the results of the spectroscopic
identifications of one of such brighter samples. The X-ray sources
have been detected by the {\it BeppoSAX}-MECS instruments in the 5-10
keV band in the framework of the High Energy LLarge Area Survey
(HELLAS).  Preliminary results have been presented in Fiore et
al. (1999) and La Franca et al. (2001). The whole survey and the
catalogue is described by Fiore et al. (2001). The data have been
analyzed in the framework of the synthesis models for the X-ray
background by Comastri et al. (2001), and the correlation with the
soft X-rays has been investigated by Vignali et al. (2001).

In section 2 we describe our X-ray and optical observations. In
section 3 we present an analysis of the evolution of AGN in the
2-10 keV band. Because of the reasons previously described, the selection
and definition of type 2/absorbed sources is still not clear, and
thus we restricted our evolutionary studies to type 1 AGN only. The
results are discussed in section 4. 

\placefigure{f1}

\section{The spectroscopic identifications}

The spectroscopic follow up of the HELLAS sources has been carried out
in a subsample enclosed in a region with $\delta < 79^\circ$, and
outside $5^h<\alpha<6.5^h$ and $17^h<\alpha<20^h$. In this region the
number of sources is 118 out of a total of 147. Their flux
distribution is shown in Figure 1 and the sky coverage is shown in
Figure 2 and listed in Table 1.

The {\it BeppoSAX} X-ray positions have an uncertainty of about 1-1.5
arcmin, the larger at larger off-axis distances. We have thus searched
for optical counterparts having R magnitude brighter than 21.0 in a
circular region of 1-1.5 arcmin of radius around the HELLAS positions
(see below and section 3.1.1 for a discussion on the choice of this
optical limit). In the case of large off-axis distances, the larger
error-boxes (1.5$'$) have been used. 25 sources have been identified
with cross-correlation with existing catalogues (NED), and 49 have
been investigated at the telescope. The total resulting sample of 74
sources has been built up in such a way that to a) randomly sample the
flux distribution of the parent catalogue of 118 sources (see Figure
1), and b) to avoid possible biases introduced by the
cross-correlation with existing catalogues (see appendix A for a
detailed discussion).

%
%
\setcounter{figure}{1}
\begin{inlinefigure}
\psfig{figure=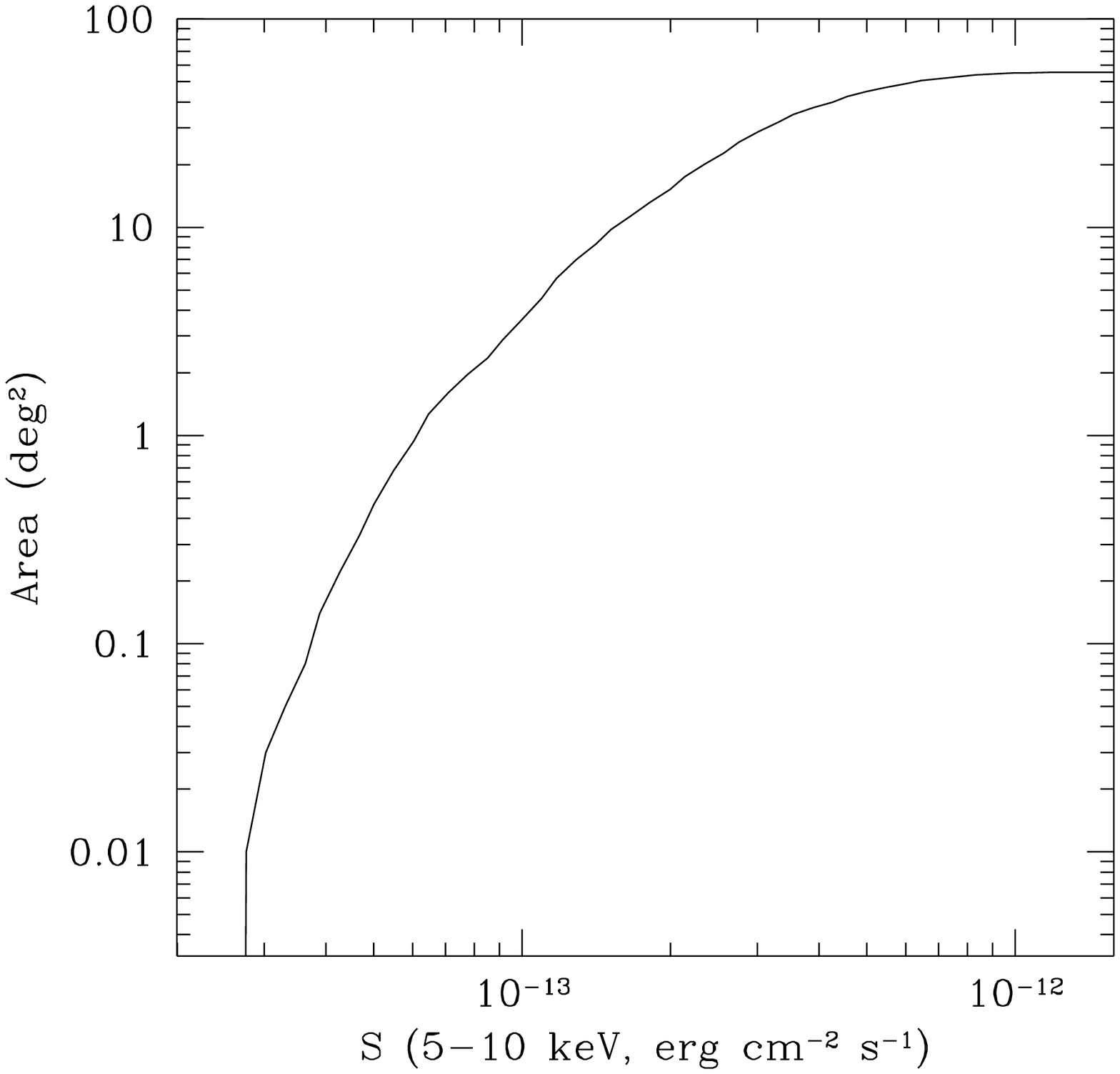,angle=0,width=8.1cm}
\caption{The sky coverage of the sample of 118 sources from HELLAS
used in this analysis.}
\figurenum{2}
\addtolength{\baselineskip}{5pt}
\end{inlinefigure}

%
%
\begin{deluxetable}{cl}
\renewcommand\baselinestretch{1.0}
\tablewidth{0pt}
%
%
%
\parskip=0.2cm
\tablenum{2}
\footnotesize
\tablecaption{Observing log}
\tablehead{
Date & Telescope}
\startdata
1998 Dec 28: 3 nights & ESO/3.6m EFOSC2    \cr 
1999 Jun \phn8: 3 nights & ESO/3.6m EFOSC2    \cr
2000 Jan \phn3: 3 nights & ESO/3.6m EFOSC2    \cr
2000 Jul\phn 29: 3 nights & ESO/3.6m EFOSC2    \cr
\enddata
\label{tab2}
\end{deluxetable}

The follow-up observations have been carried out using the EFOSC2 at
the 3.6m ESO telescope. Previously published identifications (Fiore et
al. 1999) have been carried out with the RC spectrograph (RCSP) at the
Kitt Peak 4m telescope, the FAST spectrograph at the Whipple 60$''$
telescope, and the Hawaii 88$''$ telescope. Long slit and multi object
spectroscopy have been carried with resolution between 7 and 16\AA\
. The reduction process used standard MIDAS facilities (Banse et
al. 1983). The raw data were sky-subtracted and corrected for
pixel-to-pixel variations by division with a suitably normalized
exposure of the spectrum of an incandescent source (flat-field).
Wavelength calibrations were carried out by comparison with exposures
of He-Ar, He, Ar and Ne lamps. Relative flux calibration was carried
out by observations of spectrophotometric standard stars (Oke 1990).
The campaign dates of the observations carried out at ESO are listed
in Table 2.

We have divided the identified sources in AGN1, AGN2, galaxies with
narrow emission lines showing moderate-to-high degree of excitation
(ELG), BL Lac objects, clusters of galaxies and stars. AGN2 includes
AGN1.5, AGN1.8 and AGN1.9. No distinction has been made between
radio-loud and radio quite objects. The radio properties of the HELLAS
sources will be discussed in a forthcoming paper (P. Ciliegi, in
preparation). The R-band magnitudes of these identifications have been
pushed down to the limit of obtaining about less than 10$\%$
probability of having a spurious identification, due to chance
coincidences, for each of the HELLAS sources (i.e. 90$\%$ reliability
for the whole sample). For AGN1 we have chosen a limit of R=21 where
the surface density is $\sim$60-70/deg$^2$ (Zitelli et
al. 1992). Tresse et al. (1996) found that 17$\%$ of galaxies with
$z<0.3$ host an emission line spectrum not typical of excitation from
starburst (but typical of AGN2 activity), this percentage reduces to
at minimum 8\% if the maximum likely effect of absorption under the
Balmer lines is taken into account. They also found that the galaxies
having forbidden emission lines without being AGN2 are at least
36$\%$. These objects are mainly starburst galaxies and low ionization
narrow emission line galaxies (LINER). We will call here all these
sources simply emission line galaxies (ELG). We have assumed
roughly constant these percentages and thus chosen a limit of R=19 for
AGN2 where the surface density of all galaxies is about 500/deg$^2$
and consequently the expected density of AGN2 is $\sim$70-80/deg$^2$.
For ELG we have chosen a limit of R=17.5 where the surface density of
all galaxies is about 160/deg$^2$, and therefore the expected density
of ELG is $\sim$60/deg$^2$.

The classification of the narrow emission line galaxies has been
carried out using standard line ratio diagnostics (Tresse et
al. 1996). Some of the spectra showed H$_\alpha$ in emission and
H$_\beta$ having a) very small emission, or b) absent, or c) small
absorption. Such spectra show however spectral features typical of
strong-emission-line spectra such as [OII]$\lambda 3727$,
[OIII]$\lambda 4959,5007$, and [SII]$\lambda 6725$. In such cases the
[OIII]/H$_\beta$ push the diagnostic ratio in the locus of AGN2, but
stellar absorption at H$_\beta$ could be significant affecting the
classification. Following Tresse et al. (1996), we have tested the
classification under the assumption of a possible absorption of
EW=2\AA\ over H$_\beta$.

We have spectroscopically identified 36 out of the 49 HELLAS sources
whose field has been investigated at the telescopes. For 13 sources we
have found no optical counterparts brighter than the chosen magnitude
limits (see discussion above).  The astrometric accuracy of the new
generation X-ray telescopes ({\it Chandra} and {\it XMM-Newton}) would
probably allow an unambiguous identification of the counterparts of
these HELLAS sources, but in the framework of this work these fields
have been declared {\it empty}. These sources have harder X-ray
spectra than the total sample. They have an average softness ratio of
$<{{S-H}\over{S+H}}>=-0.5\pm0.3$ in comparison with the average value
$-0.1\pm0.4$ of the total sample of 118 sources, where S and H are the
1.3-4.5 keV the 4.5-10 keV counts, respectively (see Fiore et
al. 2001). The absence of bright counterparts and the average harder
X-ray spectrum favor the hypothesis that most of these sources
preferably harbor absorbed AGN. If these sources correspond to optical
''normal'' galaxies, such as those observed by {\it Chandra} and {\it
XMM-Newton} (see e.g. Fiore et al. 2000; Barger et al. 2001;
Tozzi et al. 2001; Hornschemeier et al. 2001;
Comastri et al. 2002), we would have not been able to properly identify them
inside our error-boxes. However, our sample is statistically well
defined, and from the number of empty fields we can directly derive an
upper limit of $\sim$18\% (13/74) for the presence of these optical
normal galaxies in our survey.

The list of the observed HELLAS sources with their most probable
spectroscopic identification is shown in Table 3. 80\% of our
identifications are inside an error-box of 60 arcsecs.  2 bright
galaxies (R$\sim$15-16) are at a distance of about 110$''$.  The 13
empty fields are listed in Table 4. The optical spectra are shown in
Figure 3; as a reference, a list of the most typical emission lines for
AGN are over-plotted with the corresponding redshift. In Table 5 the
EW, FWHM, FWZI in the observed frame of the most relevant emission
lines are listed.

In total, 63$\%$ (74/118) of our HELLAS subsample has been searched
for spectroscopic identification. 61 have been identified: 37 AGN1, 9
AGN2, 5 ELG, 6 Clusters, 2 BL Lac, 1 Radio Galaxy and 1 Star.

\begin{deluxetable}{ccrrrllll}
\renewcommand\baselinestretch{1.0}
%
%
%
\tablewidth{0pt}
\parskip=0.2cm
\tablenum{3}
\footnotesize
\tablecaption{Identified Hard X-ray Sources in the HELLAS Catalogue}
\tablehead{
{\it BeppoSAX} position & Optical position & & &F$_X$ \\
RA (2000) DEC  & RA (2000) DEC & $\Delta_{xo}$ & Off$_x$ & 5-10~keV & R & Spectr. ID   & $z$\\
 $h~~m~~s~~~~~~~^{\circ}~~~'~~~''~$ & $h~~m~~s~~~~~~~^{\circ}~~~'~~~''~$ & $''$ &$'$&(1)&&
}
\startdata
00 26 36.5 $-$19 44  14 & 00 26 36.1 $-$19 44  16 &   7 &  13&   3.4 & 18.1 & AGN2  &  0.238  \\ 
00 27 09.9 $-$19 26  31 & 00 27 09.8 $-$19 26  14 &  18 &   6&   1.8 & 17.7 & AGN1  &  0.227$^a$ \\ 
00 45 49.6 $-$25 15  14 & 00 45 46.3 $-$25 15  50 &  59 &  24&   3.3 & 17.5 & AGN2  &  0.111  \\ 
01 21 56.8 $-$58 44  05 & 01 21 56.9 $-$58 44  42 &  37 &  16&   2.6 & 16.8 & AGN2  &  0.118 \\ 
01 34 33.3 $-$29 58  38 & 01 34 33.8 $-$29 58  16 &  24 &   6&   0.9 & 18.0 & AGN1  &  2.217 \\ 
01 35 30.2 $-$29 51  22 & 01 35 32.7 $-$29 52  02 &  53 &   8&   0.9 & 17.7 & AGN1  &  1.344 \\ 
01 40 08.9 $-$67 48  13 & 01 40 14.7 $-$67 48  54 &  53 &   8&   2.8 & 12.4 & Star  &  0.000 \\ 
02 42 01.8 $+$00 00  46 & 02 42 00.9 $+$00 00  22 &  29 &     10&   1.5 & 18.6 & AGN1  &  1.112$\dag$ \\ 
03 15 45.0 $-$55 29  26 & 03 15 47.5 $-$55 29  04 &  31 &     15&   2.7 & 17.9 & AGN1  &  0.464$\dag$ \\ 
03 17 32.4 $-$55 20  12 & 03 17 32.7 $-$55 20  26 &  14 &     21&   4.1 & 17.5 & AGN1  &  0.406$\dag$ \\ 
03 33 09.6 $-$36 19  40 & 03 33 12.2 $-$36 19  48 &  33 &     15&   4.0 & 17.5 & BLac  &  0.308$\dag$ \\ 
03 34 07.4 $-$36 04  22 & 03 34 07.5 $-$36 04  04 &  19 &   5&   1.9 & 20.1 & AGN1  &  0.904 \\ 
03 36 51.3 $-$36 15  57 & 03 36 54.0 $-$36 16  06 &  34 &     15&   3.7 & 17.7 & AGN1  &  1.537$\dag$ \\ 
04 37 14.5 $-$47 30  58 & 04 37 11.8 $-$47 31  48 &  57 &  16&   2.7 & 17.3 & AGN1  &  0.142 \\ 
04 38 47.9 $-$47 29  07 & 04 38 47.0 $-$47 28  02 &  67 &  20&   4.7 & 20.5 & AGN1  &  1.453 \\ 
06 46 39.3 $-$44 15  35 & 06 46 37.6 $-$44 15  34 &  19 &  17&   4.3 & 16.6 & AGN1  &  0.153 \\ 
07 21 29.6 $+$71 14  04 & 07 21 36.2 $+$71 13  24 &  52 &      8&   0.8 & 17.7 & AGN1  &  0.232$\dag$ \\ 
07 41 40.3 $+$74 14  58 & 07 41 50.0 $+$74 14  48 &  41 &     23&  30.7 & .... & Clust.&  0.216$\dag$ \\ 
07 43 09.1 $+$74 29  20 & 07 43 12.6 $+$74 29  36 &  22 &      7&   6.0 & 16.4 & AGN1  &  0.312$\dag$ \\ 
08 37 37.2 $+$25 47  49 & 08 37 37.1 $+$25 47  52 &   4 &  12&   2.6 & 16.9 & AGN1  &  0.077 \\ 
08 38 59.9 $+$26 08  14 & 08 38 59.2 $+$26 08  14 &  10 &  23&  16.4 & 15.3 & ELG   &  0.048 \\ 
10 32 15.8 $+$50 51  04 & 10 32 16.2 $+$50 51  20 &  18 &  10&   3.1 & 15.9 & AGN1  &  0.174 \\ 
10 54 19.8 $+$57 25  09 & 10 54 21.2 $+$57 25  44 &  38 &     13&   2.6 & 18.5 & AGN2  &  0.205$\dag$ \\ 
11 01 46.4 $+$72 26  11 & 11 01 48.8 $+$72 25  38 &  36 &     22&   7.3 & 16.7 & AGN1  &  1.460$\dag$ \\ 
11 02 37.2 $+$72 46  38 & 11 02 36.8 $+$72 46  40 &   3 &     21&   7.9 & 15.1 & AGN1  &  0.089$\dag$ \\ 
11 06 14.0 $+$72 43  16 & 11 06 16.1 $+$72 44  14 &  59 &      9&   1.7 & 18.5 & AGN1  &  0.680$\dag$ \\ 
11 18 11.9 $+$40 28  33 & 11 18 13.8 $+$40 28  38 &  23 &   4&   0.8 & 18.7 & AGN1  &  0.387 \\ 
11 18 46.2 $+$40 27  39 & 11 18 48.7 $+$40 26  48 &  59 &   5&   1.4 & 18.5 & AGN1  &  1.129 \\
12 04 07.6 $+$28 08  31 & 12 04 04.0 $+$28 07  24 &  82 &     16&   5.1 & .... & Clust.&  0.167$\dag$ \\ 
12 17 50.3 $+$30 07  08 & 12 17 52.1 $+$30 07  02 &  25 &     20&   3.5 & 14.0 & BLLac &  0.237$\dag$ \\ 
12 18 55.0 $+$29 58  12 & 12 18 52.5 $+$29 59  02 &  59 &  13&   2.0 & 18.6 & AGN2  &  0.176 \\ 
12 19 45.7 $+$47 20  43 & 12 19 52.3 $+$47 20  58 &  69 &      8&   1.2 & 19.3 & AGN1  &  0.654$\dag$ \\ 
12 22 06.8 $+$75 26  17 & 12 22 07.0 $+$75 26  18 &   2 &      7&   2.5 & .... & Clust.&  0.240$\dag$ \\ 
12 40 26.0 $-$05 13  20 & 12 40 27.8 $-$05 14  02 &  50 &  12&   3.1 & 18.8 & AGN1  &  0.300 \\ 
12 40 29.6 $-$05 07  46 & 12 40 36.4 $-$05 07  52 & 102 &  17&   1.9 & 15.2 & ELG   &  0.008 \\ 
13 04 38.2 $-$10 15  47 & 13 04 35.6 $-$10 15  48 &  39 &   6&   1.4 & 20.1 & AGN1  &  2.386 \\ 
13 05 32.3 $-$10 32  36 & 13 05 33.0 $-$10 33  20 &  45 &     22&  19.3 & 14.9 & AGN1  &  0.278$\dag$ \\ 
13 36 34.3 $-$33 57  48 & 13 36 39.0 $-$33 57  58 &  60 &     22&   3.2 & 10.5 &RadioG.&  0.013$\dag$ \\ 
13 42 59.3 $+$00 01  38 & 13 42 56.5 $+$00 00  58 &  59 &     21&   3.2 & 18.7 & AGN1  &  0.804$\dag$ \\ 
13 48 20.8 $-$30 11  06 & 13 48 19.5 $-$30 11  56 &  52 &  14&   2.2 & 15.3 & AGN2  &  0.128 \\ 
13 48 45.4 $-$30 29  37 & 13 48 44.7 $-$30 29  46 &  13 &  14&   5.1 & 17.1 & AGN1  &  0.330 \\ 
13 50 09.4 $-$30 19  55 & 13 50 15.4 $-$30 20  10 &  80 &  11&   5.1 & 16.5 & ELG   &  0.074 \\ 
13 53 54.6 $+$18 20  33 & 13 53 54.4 $+$18 20  16 &  18 &  18&   6.8 & 17.3 & AGN1  &  0.217 \\ 
14 17 12.5 $+$24 59  29 & 14 17 18.8 $+$24 59  30 &  86 &     13&   0.7 & 19.5 & AGN1  &  1.057$\dag$ \\ 
14 18 31.1 $+$25 11  07 & 14 18 31.2 $+$25 10  50 &  17 &      9&   6.1 & .... & Clust.&  0.240$\dag$ \\ 
15 19 39.9 $+$65 35  46 & 15 19 33.7 $+$65 35  58 &  41 &  14&   9.4 & 14.4 & AGN2  &  0.044 \\ 
15 28 47.3 $+$19 39  10 & 15 28 47.7 $+$19 38  54 &  18 &   5&   1.6 & 20.3 & AGN1  &  0.657$^b$ \\ 
16 26 59.9 $+$55 28  21 & 16 26 59.0 $+$55 27  24 &  57 &     11&  12.1 & .... & Clust.&  0.130$\dag$ \\ 
16 34 11.8 $+$59 45  29 & 16 34 18.5 $+$59 45  44 &  53 &   3&   0.8 & 19.0 & AGN2  &  0.341$^c$ \\ 
16 49 57.9 $+$04 53  33 & 16 50 00.0 $+$04 54  00 &  42 &     20&   9.5 & .... & Clust.&  0.154$\dag$ \\ 
16 50 40.1 $+$04 37  17 & 16 50 42.7 $+$04 36  18 &  71 &  25&  12.2 & 14.6 & AGN2  &  0.031 \\ 
16 52 38.0 $+$02 22  18 & 16 52 37.5 $+$02 22  06 &  14 &   5&   0.7 & 20.7 & AGN1  &  0.395 \\ 
20 42 47.6 $-$10 38  31 & 20 42 53.0 $-$10 38  26 &  80 &  21&   5.3 & 17.9 & AGN1  &  0.363 \\ 
20 44 34.8 $-$10 27  34 & 20 44 34.8 $-$10 28  08 &  35 &  15&   2.0 & 17.7 & AGN1  &  2.755 \\ 
22 26 30.3 $+$21 11  57 & 22 26 31.5 $+$21 11  38 &  26 &  14&   3.9 & 17.6 & AGN1  &  0.260 \\ 
23 15 36.4 $-$59 03  40 & 23 15 46.8 $-$59 03  14 &  85 &      8&   2.0 & 11.2 & ELG   &  0.044$\dag$ \\ 
23 19 22.1 $-$42 41  50 & 23 19 32.0 $-$42 42  28 & 116 &  21&   5.7 & 16.5 &ELG$^d$ & 0.101 \\ 
23 27 28.7 $+$08 49  30 & 23 27 28.7 $+$08 49  26 &   4 &   7&   0.5 & 18.5 &AGN1$^e$&  0.154 \\ 
23 29 02.4 $+$08 34  39 & 23 29 05.8 $+$08 34  16 &  56 &  20&   2.9 & 20.3 & AGN1  &  0.953 \\ 
23 31 55.6 $+$19 38  34 & 23 31 54.3 $+$19 38  36 &  19 &  17&   3.7 & 18.8 & AGN1  &  0.475 \\ 
23 55 53.3 $+$28 36  06 & 23 55 54.3 $+$28 35  58 &  16 &     12&   4.2 & 17.9 & AGN1  &  0.731$\dag$ \\ 
\enddata
\label{tab3}
\tablecomments{
(1) 10$^{-13}$ \ecs; 
$^\dag$ From cross-correlation with existing catalogues;
$a)$ From Colafrancesco et al. (2000);
$b)$ From Gorosobel et al. (1998), misidentified in Fiore et al. 2000;
$c)$ Wrong coordinates and magnitudes in Fiore et al. (1999);
$d)$ Also possible AGN1: H$_\beta$ has a rest frame FWHM$\sim$1000 Km/s;
$e)$ Quite noisy spectrum.  H$_\beta$ has a rest frame FWHM$\sim$1500 Km/s.
}
\end{deluxetable}

\begin{deluxetable}{cc}
\renewcommand\baselinestretch{1.0}
\tablewidth{0pt}
%
%
%
\parskip=0.2cm
\tablenum{4}
\footnotesize
\tablecaption{Empty fields in the HELLAS Catalogue}
\tablehead{
{\it BeppoSAX} position &F$_X$ \\
RA (2000) DEC  & 5-10~keV \\
$h~~m~~s~~~~~~~^{\circ}~~~'~~~''~$ & (10$^{-13}$ \ecs) 
}
\startdata
 01 34 28.6  $-$30 06 04 & 1.33\\
 01 34 49.6  $-$30 02 34 & 0.71\\ 
 09 46 05.3  $-$14 02 59 & 2.82\\
 09 46 32.8  $-$14 06 16 & 3.22\\ 
 13 04 24.3  $-$10 23 53 & 1.28\\ 
 13 04 45.1  $-$05 33 37 & 1.26\\
 13 48 24.3  $-$30 25 47 & 3.15\\
 14 11 58.7  $-$03 07 02 & 3.93\\ 
 22 03 00.5  $-$32 04 18 & 2.83\\ 
 22 31 49.6  $+$11 32 08 & 1.94\\
 23 02 30.1  $+$08 37 06 & 2.67\\
 23 02 36.2  $+$08 56 42 & 3.17\\ 
 23 16 09.8  $-$59 11 24 & 1.32\\ 

\enddata
\label{tab4}
\tablecomments{Within 1 arcmin there are no AGN1 down to R=21,
no AGN2 down to R=19 and no ELG down to R=17.5.
}
\end{deluxetable}

\placefigure{f3}

\placetable{tab1}
%
%
\placefigure{f4}

\begin{inlinefigure}
\figurenum{4}
\psfig{figure=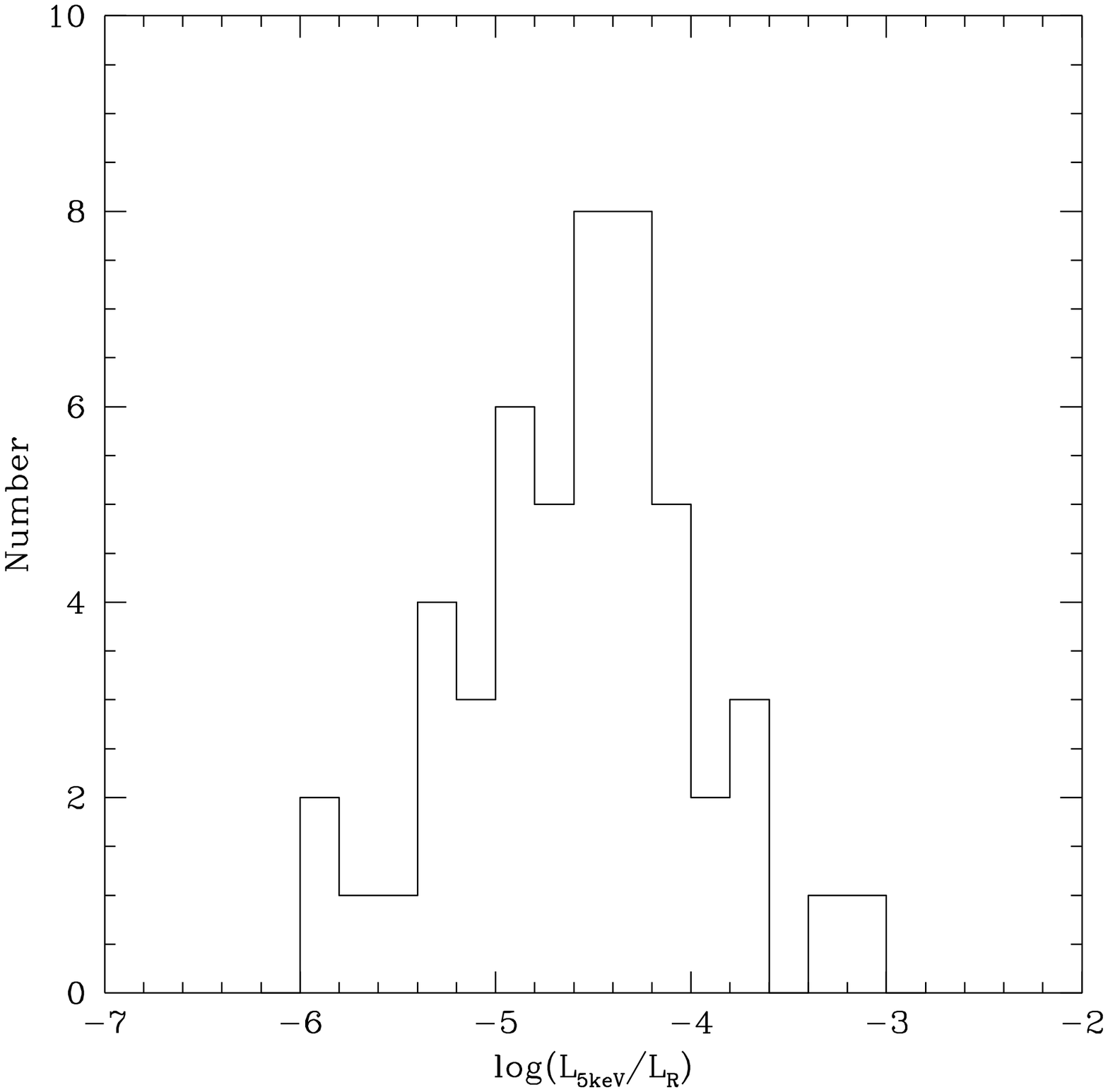,angle=0,width=8.2cm}
\caption{
The histogram of the Log(L$_{5~keV}$/L$_{R}$) ratio for a sample of 50 (25
X-ray selected and 25 optically selected) AGN1 (see text).  We fitted
the distribution with a Gaussian having mean -4.58 and standard
deviation 0.61. This distribution has been assumed as representative
of the whole population of type 1 AGNs at any redshift and luminosity,
and used to compute the fraction of AGN1 fainter that R=21 missed
from the optical spectroscopic identifications of the HELLAS sample. 
}
\addtolength{\baselineskip}{30pt}
\end{inlinefigure}

\section{Analysis on the evolution of AGN1}

Because of the difficulties previously described in quantifying all
the selection biases in building up samples of type 2 (and/or
absorbed) AGN, we have decided to limit our evolutionary analysis to
AGN1 only.  See appendix A for a discussion on the absence of
substantial biases on our sample of AGN1.  We have assumed H$_0$=50
km/s/Mpc and either the ($\Omega_m$,$\Omega_\lambda$)=(1.0,0.0) or the
($\Omega_m$,$\Omega_\lambda$)=(0.3,0.7) cosmologies. We have computed
the 2-10 keV luminosity for AGN1 assuming a typical spectral energy
distribution as computed by Pompilio, La Franca, \& Matt (2000), which
assumes a power slope in energy with index $\alpha$=0.9 (${dF\over
dE}\propto E^{-\alpha}$) and takes into account the reflection. This
spectrum is roughly approximated by a single power $\alpha$=0.6 in the
range 2-50 keV (i.e. up to $z$=4). This is in agreement with the
average slope of the spectra of AGN1 from the HELLAS sample (Vignali
et al. 2001).

\subsection{The data}

\subsubsection{The completeness of the sample of AGN1 from HELLAS}

As our spectroscopic identifications are limited to magnitudes
brighter than $R=21$, in order to estimate the density of AGN1, we
have evaluated which is their completeness in our sample. We have
assumed that all AGN1 follow the same distribution of the ratio of the
5 keV to R-band optical luminosity Log(L$_{5~keV}$/L$_{R}$) without
dependencies on luminosity and/or redshift. We have estimated the
relationship between optical and hard X-ray luminosity for AGN1 as
follow. We selected a mixed sample of both 25 optical and 25 hard
X-ray selected type 1 AGN, mostly in the redshift range 0$<$$z$$<$2,
from Mineo et al. (2000), George et al. (2000) and Akiyama et
al. (2000). The 25 optically selected AGN1 have an average
Log(L$_{5~keV}$/L$_{R}$) ratio of -4.78 with a standard deviation of
0.59, while the 25 X-ray selected AGN1 have an average
Log(L$_{5~keV}$/L$_{R}$) ratio of -4.38 with a standard deviation of
0.58.  In Figure 4 the histogram of the ratio of 5 keV and R-band
optical luminosity of the total sample of 50 AGN1 at $z$=0 is
shown. We fitted the distribution with a Gaussian having mean -4.58
and standard deviation 0.61. We have estimated that with this
distribution, a limit of $R=21$ corresponds to only 1 AGN1 lost from
our HELLAS survey, and consequently we can say that a limit of $R=21$
assure a high level of completeness for the identification of AGN1 at
the fluxes of our HELLAS survey. At variance, as already discussed,
our HELLAS identification programme is biased against absorbed type 2
sources which probably harbor in the 13 empty fields found.

Although the incompleteness for AGN1 in the HELLAS sample is quite
low, during the computation of the LF of AGN1, we have estimated the
correct area coverage for our survey by multiplying the area coverage
corresponding to each point of the $L_X/z$ plane for: a) the
spectroscopic completeness of the HELLAS sample (74/118), and b) the
fraction of found AGN1 as a function of $z$, as described before.

\subsubsection{The other samples}

In order to increase the statistical significance of our analysis, the
HELLAS data have been combined to other hard X-ray samples of AGN1
identified by Grossan (1992), Boyle et al. (1998), and Akiyama et
al. (2000).

The sample of Grossan (1992) consists of 84 AGN1 and 12 AGN2 detected
by {\it HEAO1}, predominantly at low redshift $z<0.3$. The sample
covers an area of 26919 deg$^2$ down to a flux limit of $1.8\times
10^{-11}$ \ecs. The quoted fluxes at 5 keV have been converted to
fluxes in the 2-10 keV assuming $\alpha=0.6$.

The sample of Boyle et al. (1998) consists of 12 AGN1 and 6 AGN2
detected by {\it ASCA}. 2-10 keV counts have been converted to fluxes
with a count-rate-to-flux conversion factor of
$5.8\times10^{-11}$ \ecs/count, assuming $\alpha=0.6$ (Georgantopoulos
et al., 1997). The sky coverage has been taken from Georgantopoulos et
al.\ (1997, see their figure 1).

The sample of Akiyama et al.\ (2000) consists of 25 AGN1, 5 AGN2, 2
clusters, 1 star and 1 unidentified source.  The sources have been
detected by {\it ASCA}. In order to convert the counts of the sky
coverage presented in their Table 1, a conversion has been applied
assuming that 1 count corresponds to $0.9\times 10^{-13}$ \ecs in the
2-10 keV band for an X-ray source with a slope $\alpha$=0.6.

A total of 158 AGN1 have been used in the redshift range
0$<$$z$$<$3. Their luminosity-redshift distribution is shown in Figure
5.  In this distribution a gap between the sample of Grossan from {\it
HEA01} and the other fainter samples from {\it ASCA} and {\it
BeppoSAX} is evident. This is because surveys at fluxes in the range
$10^{-12}$--$10^{-11}$ \ecs in the 2-10 keV energy band where some
hundreds of deg$^2$ need to be covered are still missing. These
surveys are in program with {\it XMM-Newton} (see e.g. Barcons et al.
2001).

\subsection{The method}

The best-fit parameters for the 2-10 keV LF and its cosmological
evolution have been derived by minimization of the $\chi^2$ computed
by comparison of the observed and expected number of AGN1 in
6$\times$3 bins in the $L_X/z$ plane. The adopted luminosity/redshift
grid was: 6 Log$L_X$ bins, with $\Delta$Log$L_X$=1 in the range
42.2-48.2, and 3 redshift bins (0.0-0.2, 0.2-1.0, 1.0-3.0).  Table 6
summarizes the results for a number of different models. The errors
quoted for the parameters are 68$\%$ (1$\sigma$) confidence
intervals. They correspond to variations of $\Delta\chi^2$=1.0,
obtained perturbing each parameter in turn with respect to its
best-fit value, and looking for a minimization with the remaining
parameters free to float. The fitted LF have been tested with the 2D
Kolmogorov-Smirnov (2DKS) test.  Two different functional forms have
been used. Following Ceballos and Barcons (1996) and Boyle et
al. (1998), we used the pure luminosity evolution (PLE) for the AGN1
LF. We also used the luminosity dependent density evolution (LDDE)
model similar to the one fitted in the soft X-rays by Miyaji et
al. (2000).

The local ($z$=0) AGN1 LF used for the PLE model has been represented by a
two-power-law:
\begin{eqnarray}
	\begin{array}{ll}

{d\Phi(L_{X}) \over dL_{X}} = A (L_X^{*(\gamma_1-\gamma_2)}) L_{X}^{-\gamma _1} & : L_X \leq L_X^*(z=0), \nonumber \\
{d\Phi(L_{X}) \over dL_{X}} = A L_{X}^{-\gamma _2} & : L_X > L_X^*(z=0),\nonumber 
	\end{array}
\end{eqnarray}

\noindent
where $L_{X}$
is expressed in units of $10^{44}{\rm erg\,s^{-1}}$.
A standard power-law luminosity evolution model was
used to parameterize the cosmological evolution of this LF:
$ L_X^*(z) = L_X^*(0)(1+z)^{k}$.

%
%
\placefigure{f5}

\begin{inlinefigure}
\figurenum{5}
\psfig{figure=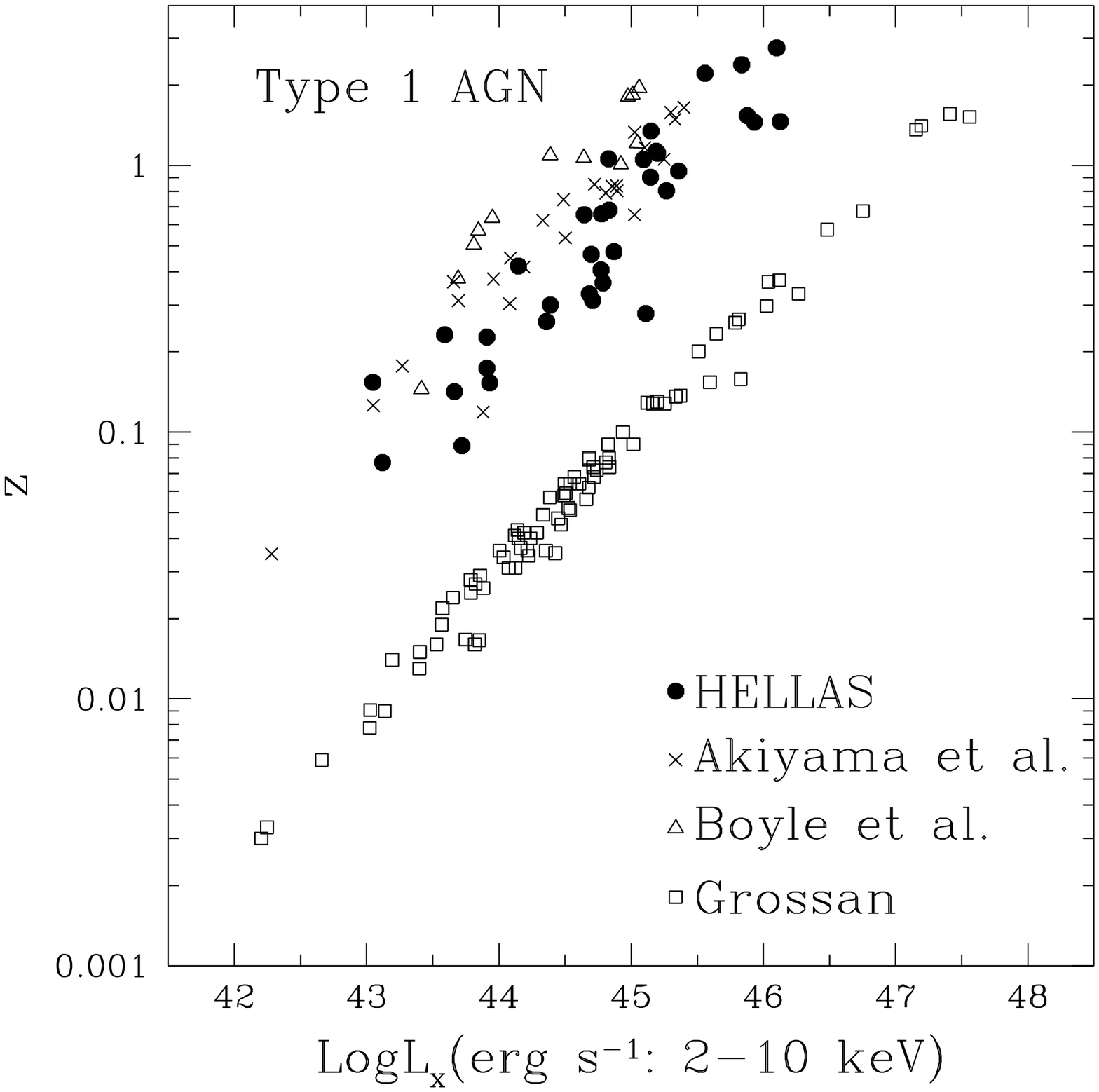,angle=0,width=8.1cm}
\caption{
The luminosity/redshift distribution for all the 158 AGN1 used to
compute the hard X-ray 2-10 keV luminosity function. H$_0$=50 km/s/Mpc
and ($\Omega_m$,$\Omega_\Lambda$)=(1.0,0.0) have been assumed.  It
includes 37 AGN1 from HELLAS, 84 AGN1 from Grossan (1992), 12 from
Boyle et al. (1998) and 25 from Akiyama et al. (2000).  
}
\addtolength{\baselineskip}{10pt}
\end{inlinefigure}

In the case of the LDDE model, as an analytical expression of the
present day ($z$=0) luminosity function, we used the smoothly
connected two power-law form:

$$ {d\Phi(L_{\rm x},z=0) \over d{\rm Log}L_{\rm x}} = A[(L_{\rm x}/L_\ast)^{\gamma_1} + (L_{\rm x}/L_\ast)^{\gamma_2}]^{-1}.$$

\noindent
The description of the evolution is given by a factor $e(L_{\rm x},z)$
such that:

$$ {d\Phi(L_{\rm x},z) \over d{\rm Log}L_{\rm x}} = {d\Phi(L_{\rm x},z=0) \over d{\rm Log}L_{\rm x}} \times e(L_{\rm x},z),$$

where
\begin{eqnarray}
  \lefteqn{e(L_{\rm x}, z) = } \nonumber \\
      & \left\{ 
	\begin{array}{ll}
	  (1+z)^{\max(0,{p1}-{\alpha}\,{\rm Log}\; 
	  {[L_{\rm a}/L_{\rm x}]})} 
	     & (z \leq z_{\rm c}; L_{\rm x}<L_{\rm a}) \\ 
	  (1+z)^{p1}
	     & (z \leq {z_{\rm c}}; L_{\rm x}\ge L_{\rm a})\nonumber \\ 
        e(L_{\rm x},{z_{\rm c}})
         \left[(1+z)/(1+{z_{\rm c}}) \right]^{p2} 
	     & (z>{z_{\rm c}}). \\
	\end{array}
       \right.
\end{eqnarray}

\begin{deluxetable}{lclllllllr}
\renewcommand\baselinestretch{1.0}
\tablewidth{0pt}
%
%
%
\parskip=0.2cm
\tablenum{6}
\footnotesize
\tablecaption{The 2-10 keV AGN LF parameters}
\tablehead{
 Model & $\Omega_m,\Omega_\lambda$ & $\gamma_1$ & $\gamma_2$ & Log$L^*$ & $k~or~p1$ & $z_{cut}$ & $A^a$ & $Prob$ & f$_{XRB}$$^1$ 
}
\startdata
0) PLE from Boyle     & 1.0,0.0 & 1.73     & 2.96     & 44.16     & 2.00     & ...     & $8.2\times 10^{-7}$      & $0.04^b$ & 24\\

1) PLE                & 1.0,0.0 & 1.83     & 3.00     & 44.17     & 2.12     & ...     & $9.8\times 10^{-7}$      & $0.22$ & 35 \\
2) PLE with $z_{cut}$ & 1.0,0.0 & 1.87     & 3.03     & 44.17     & 2.52     & 1.39    & $8.9\times 10^{-7}$      & $0.31$ & 34 \\
3) LDDE               & 1.0,0.0 & 0.68     & 2.02     & 44.03     & 4.66     & 1.55$^c$& $2.0\times 10^{-6}$ $^{d}$ & $0.47$ & 40 \\

4) PLE                & 0.3,0.7 & 1.93     & 2.97     & 44.24     & 2.22     & ...     & $8.7\times 10^{-7}$      & $0.47$ & 56 \\
5) PLE with $z_{cut}$ & 0.3,0.7 & 1.94     & 2.97     & 44.24     & 2.26     & 2.38    & $8.7\times 10^{-7}$      & $0.70$ & 54 \\
6) LDDE               & 0.3,0.7 & 0.69     & 1.95     & 43.96     & 4.56     & 1.58$^c$& $2.3\times 10^{-6}$ $^d$ & $0.77$ & 60 \\
\\
7) LDDE softX         & 1.0,0.0 & 0.62$^c$ & 2.25$^c$ & 44.13     & 5.40$^c$ & 1.55$^c$& $1.4\times 10^{-6}$ $^{c,d}$ & $0.11$ & 48 \\

\\
68$\%$ confidence errors &  & $^{+0.12}_{-0.16}$ &$^{+0.11}_{-0.09}$ &$^{+0.13}_{-0.15}$ & $^{+0.13}_{-0.14}$ &$^{+0.56}_{-0.25}$ & 8$\%$~~~~ \\
\enddata
\tablecomments{
(1) \% of the XRB: $I_{2-10}=
1.95\times 10^{-11}$ erg cm$^{-2}$ s$^{-1}$ deg$^{-2}$ (Chen,
Fabian and Gendrau 1997)\\
(a) Mpc$^{-3}$($10^{44}$erg s$^{-1})^{-1}$; 
(b) $\chi^2$ probability; 
(c) fixed; 
(d) Mpc$^{-3}($erg s$^{-1})^{-1}$.
}
\label{tab6}
\end{deluxetable}

The parameters are defined as in Miyaji et al. (2000). In particular,
$\alpha$ is proportional to the decrease of the density evolution of
the faint objects. In this parameterization of the LF $L_X$ is
expressed in units of ${\rm erg\,s^{-1}}$. In both models the LF has
been computed for luminosities brighter than LogL$_{2-10 keV}$=42.2.
Using this parameterization, for
($\Omega_m$,$\Omega_\lambda$)=(1.0,0.0), a fit was done by keeping
fixed the parameters $z_c$=$1.55$ (the redshift at which the evolution
stops), $\alpha$=$2.5$, and the ratio $L_{\rm a}/L_\ast=10^{0.42}$ to
the value found by Miyaji et al. (2000) for AGN1 only (their LDDE1
model) in the soft X-rays, and leaving all the remaining parameters
(the two slopes $\gamma_1$ and $\gamma_2$, the break luminosity
$L_\ast$, and the speed of the density evolution $p1$) free to vary.
In the case of an ($\Omega_m$,$\Omega_\lambda$)=(0.3,0.7) Universe, as
the fit for AGN1 only from Miyaji et al. (2000) was not available, their
parameters from the fit for all AGNs (AGN1+AGN2) have been used:
$z_c$=$1.58$, $\alpha$=$2.6$, $L_{\rm a}/L_\ast=10^{0.65}$.

%
%
\placefigure{f6}

\begin{inlinefigure}
\figurenum{6}
\psfig{figure=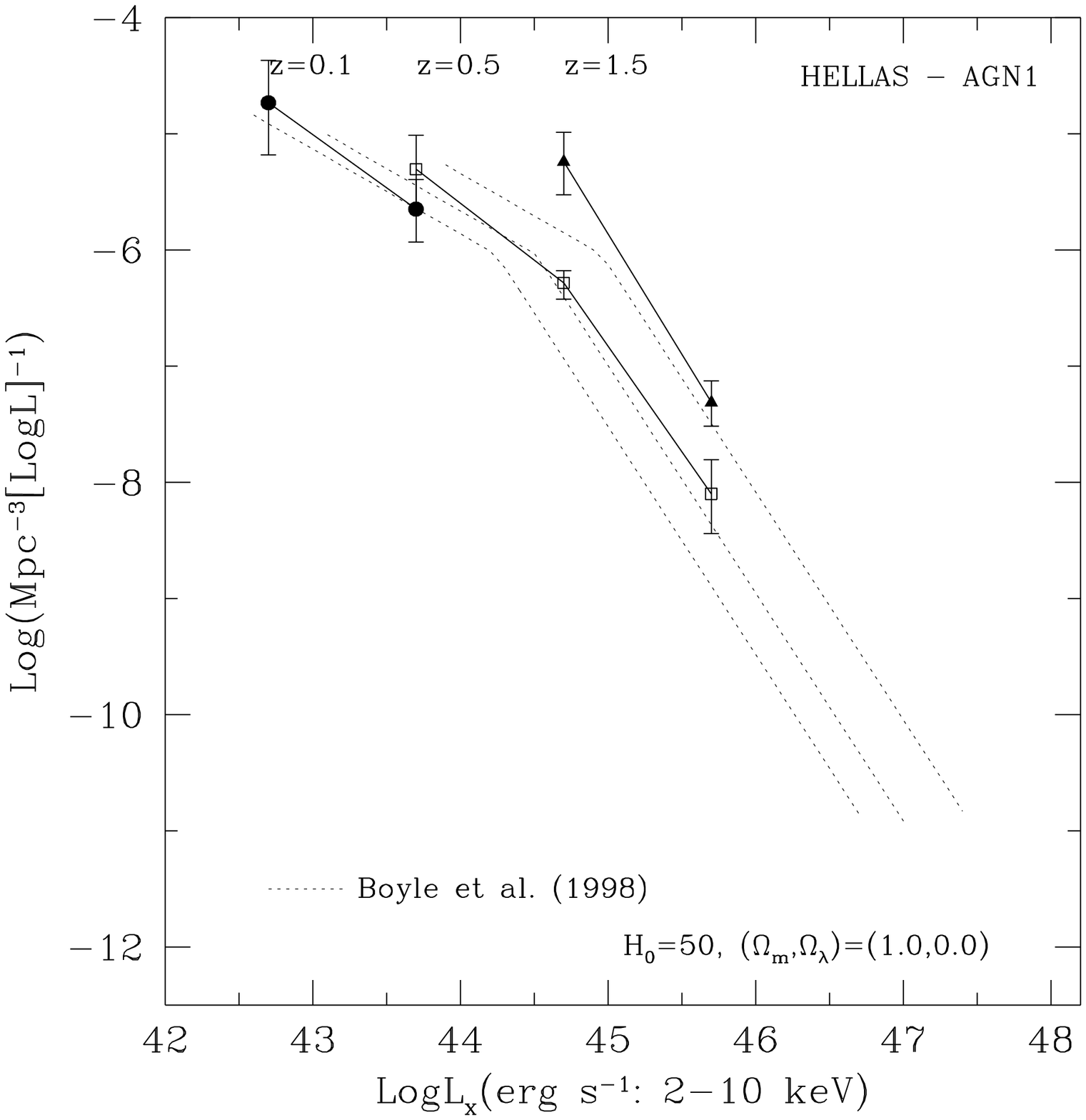,angle=0,width=8.1cm}
\caption{
The luminosity function of the 37 HELLAS AGN1. As a reference,
the dotted lines show the previously estimated luminosity function
in the 2-10 keV from Boyle et al. (1998).
}
\addtolength{\baselineskip}{5pt}
\end{inlinefigure}

\subsection{Results}

\subsubsection{The LF in the ($\Omega_m$,$\Omega_\Lambda$)=(1.0,0.0) universe}

We started our computation of the LF in an
($\Omega_m$,$\Omega_\Lambda$)=(1.0,0.0) cosmology.  Previous estimates
of the shape and evolution of the LF of AGN1 in the 2-10 keV range are
from Ceballos and Barcons (1996), and Boyle et al. (1998). Boyle et
al. (1998) combined the local sample of 84 AGN1 observed by {\it
HEAO1} from Grossan (1992) with a fainter sample of 12 AGN1 observed
by {\it ASCA} (see their distribution in the L$_X$-$z$ plane in Figure
4). They found that the 2-10 keV AGN X-ray LF is best represented by a
two-power-law function evolving according to a pure luminosity
evolution (PLE) model: $L_X(z)$$\propto$$(1+z)^{2.00}$ (see model 0 in
Table 6).

In Figure 6 the LF from only the 37 AGN1 from HELLAS in three redshift
intervals (0.0$<$$z$$<$0.2, 0.2$<$$z$$<$1.0, and 1.0$<$$z$$<$3.0) is
shown. For the sake of comparison, the best-fit LF from Boyle et
al. (1998) is also shown. The data have been represented by correcting
for evolution within the redshift bins as explained in La Franca \&
Cristiani (1997).  The error bars in the figures are based on Poisson
statistics at the 68$\%$ confidence level.

%
%
\begin{inlinefigure}
\figurenum{7}
\psfig{figure=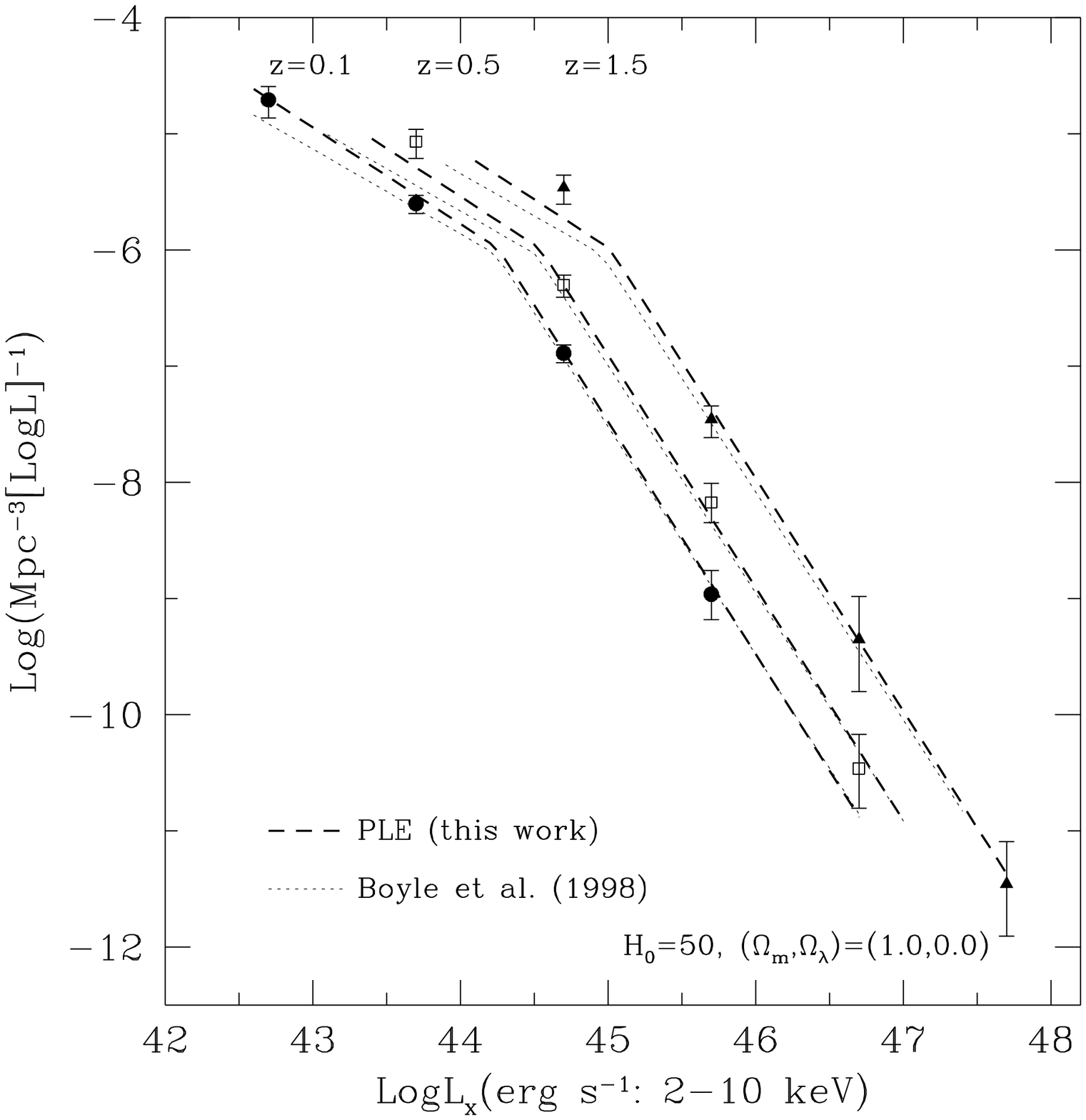,angle=0,width=8.1cm}
\caption{
The LF of the total sample of 158 AGN1 including the
37 AGN1 from HELLAS, 84 from Grossan (1992), 12 from Boyle et
al. (1998) and 25 from Akiyama et al. (2000). The excess of faint
AGN1 at high redshift is even more evident than in the previous
figure. The dotted line is the PLE fit by Boyle et al. (1998). The
dashed line is our best PLE fit (see Table 6).  }
\addtolength{\baselineskip}{5pt}
\end{inlinefigure}

The data are in rough agreement with the previous estimate of Boyle et
al. (1998), but show an excess of faint AGN1 at redshift larger than
$z$$\sim$1.0.  This feature is still present after combining our data
with the other AGN1 samples from Grossan (1992), Boyle et al. (1998),
and Akyiama et al. (2000) (Figure 7).

%
%
\begin{inlinefigure}
\figurenum{8}
\psfig{figure=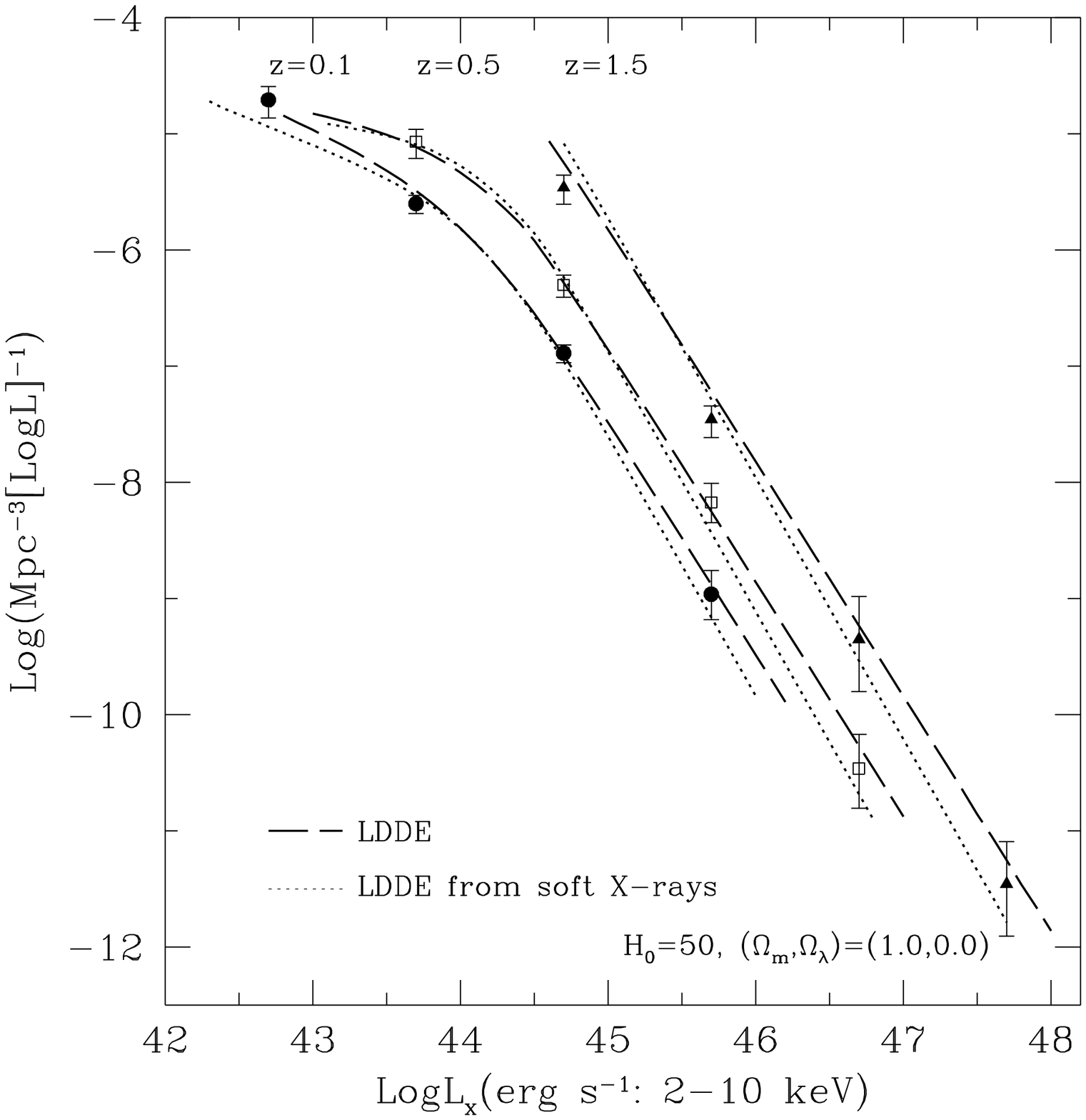,angle=0,width=8.1cm}
\caption{
The luminosity function of AGN1 fitted with the LDDE model (dashed
line).  The soft X-ray LF (Miyaji et al. 2000) is also shown (dotted
line), with the assumption LogL$_{2-10keV}$=LogL$_{0.5-2keV}$+0.33
(see text).  }
\addtolength{\baselineskip}{10pt}
\end{inlinefigure}

As already discussed, these samples all together collect 158
AGN1. These data have a $\chi^2$ probability of 0.04 to be drawn from
a PLE model such as that computed by Boyle et al. (1998) (model 0 in
Table 6). We did not use the 2DKS as this test is not appropriate in
this case: it uses the cumulative distributions in the $L_X/z$ plane
regardless of the normalization of the LF.  Our best-fit to the data
with a PLE model found a LF with similar slopes and break luminosity
as that one of Boyle et al. (1998) but with a slightly larger
evolution ($k=2.12^{+0.13}_{-0.14}$ instead of
$k$=$2.00^{+0.16}_{-0.22}$), and a significantly 20($^{+9}_{-5}$)$\%$
larger normalization (the normalization has a Poisson uncertainty of
8$\%$. See model 1 in Table 6).

\begin{inlinefigure}
\figurenum{9}
\psfig{figure=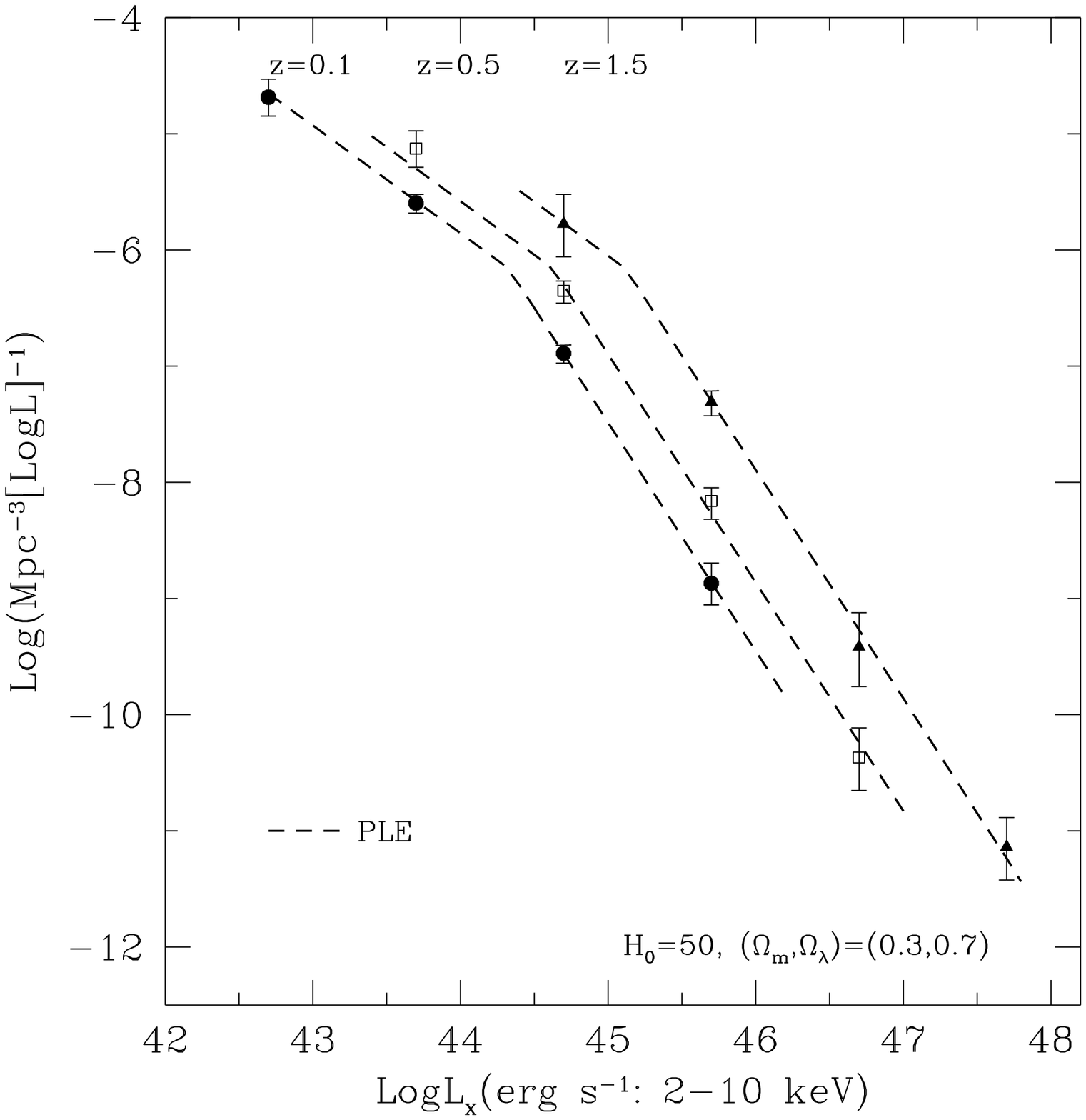,angle=0,width=8.1cm}
\caption{
The luminosity function in the ($\Omega_m$,$\Omega_\lambda$)=(0.3,0.7)
universe fitted by our PLE model (see Table 6).  }
\addtolength{\baselineskip}{10pt}
\end{inlinefigure}

%
%
\placefigure{f9}
Our larger values of the evolution parameter $k$ and normalization $A$
are originated by the necessity of better fitting the observed higher
density of faint AGN1 at high redshift. The 2DKS test gives a
probability of 0.22 for this fit (see Table 6). A even better
probability of 0.31 is obtained if a stop in the evolution is applied
at redshift $z_{cut}$=$1.39$ (model 2 in Table 6) and a larger evolution
($k$=$2.52$) is used.

%
%
\begin{inlinefigure}
\figurenum{10}
\psfig{figure=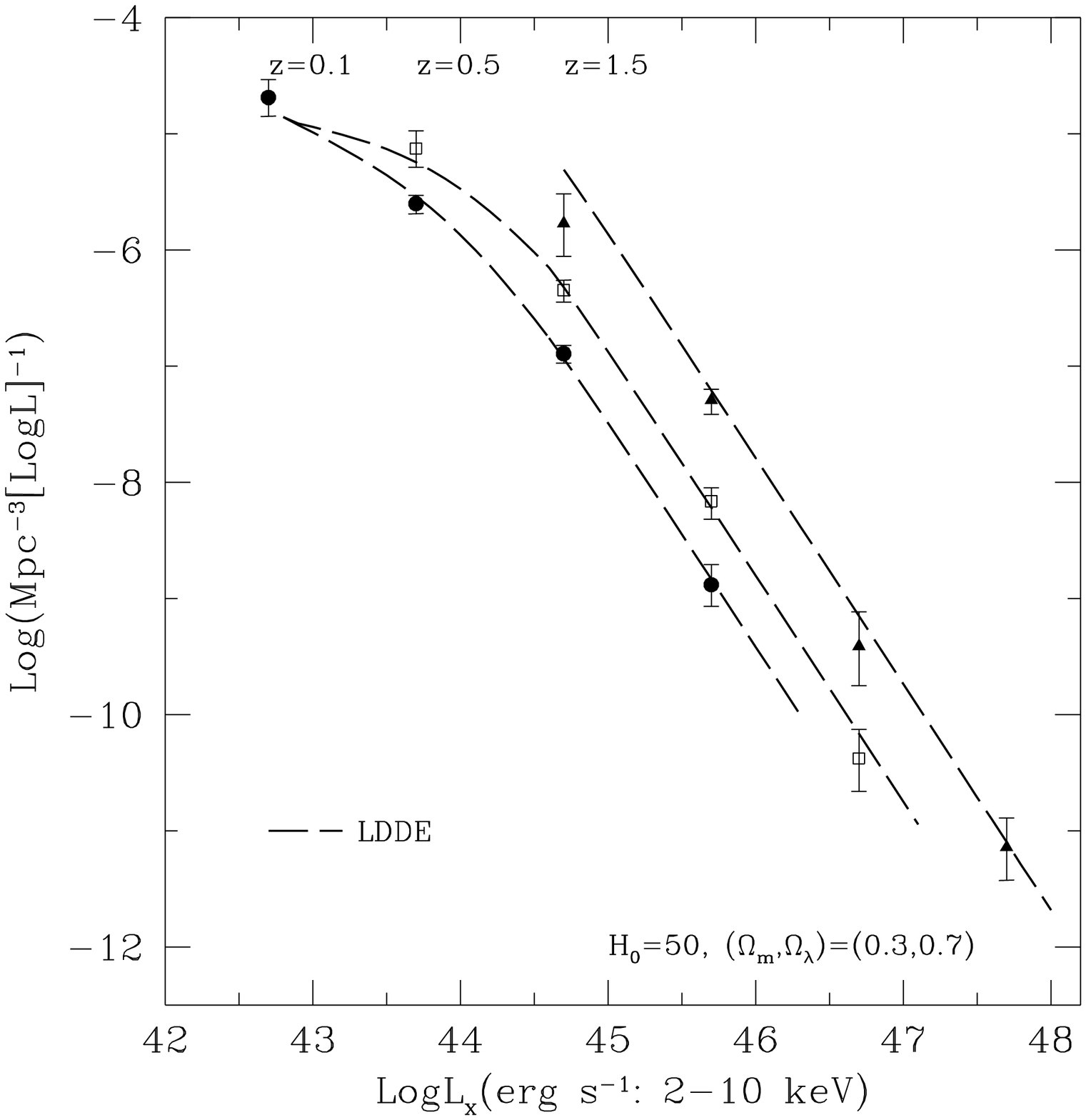,angle=0,width=8.1cm}
\caption{
The luminosity function in the ($\Omega_m$,$\Omega_\lambda$)=(0.3,0.7)
universe fitted by the LDDE model (see text and Table
6).  }
\addtolength{\baselineskip}{5pt}
\end{inlinefigure}

Although our fits are already statistically adequate, our PLE models
do not fully describe an over-density of faint AGN1 which is observed
at high redshift. The data are not sufficiently faint to properly
probe this part of the LF, but as this feature is similar to what
observed in the soft X-rays (Miyaji et al. 2000), we tried to obtain a
even better fit of the data, by using the luminosity dependent density
evolution (LDDE) model similar to the one fitted in the soft X-rays,
as described in the previous section. This model differentiates from
the PLE especially in the faint part of the LF, at luminosities lower
than $L_\ast$. In this part of the LF the density evolution decreases
in proportion to the faintness of the objects. As our data just start
to probe the part of the LF which is fainter than $L_\ast$, we cannot
expect to find directly a fit to all the parameters of the LDDE model
to our data, and we limited our analysis to a check of the
compatibility (with limited changes) of the LDDE parameters found by
Miyaji et al. (2000) to our data.  With this model we found the
best-fit to our data (model 3: LDDE, Figure 8) by keeping fixed the
parameters describing the stop in the evolution ($z_c$=$1.55$), and
the dependence on luminosity of the density evolution ($\alpha$=$2.5$)
to the value found by Miyaji et al. (2000) for AGN1 only, and leaving
all the remaining parameters free to vary (see section 3.2). The 2DKS
test gives, for this model, a probability of 0.47.

\subsubsection{The ($\Omega_m$,$\Omega_\Lambda$)=(0.3,0.7) universe}

If an ($\Omega_m$,$\Omega_\lambda$)=(0.3,0.7) cosmology is assumed,
the PLE models provide an ever better representation of the data in
comparison with what found in the
($\Omega_m$,$\Omega_\lambda$)=(1.0,0.0) Universe (see Figure 9). The
2DKS probability is 0.70 and 0.47, with and without the introduction
of the $z_{cut}$ parameter, respectively. In this case even the simple
PLE model obtain a quite good fit of the data, and the introduction of
the $z_{cut}$ parameter is necessary to stop the evolution only at
redshifts larger than 2.4. However our data contain not enough AGN1 at
redshift larger than 2 in order to obtain an accurate measure of the
$z_{cut}$ parameter (see Figure 4), therefore the errors in this
parameter are quite large. Also the LDDE model obtain a satisfactory
fit of the data (see Figure 10).

%
%
\begin{inlinefigure}
\figurenum{11}
\psfig{figure=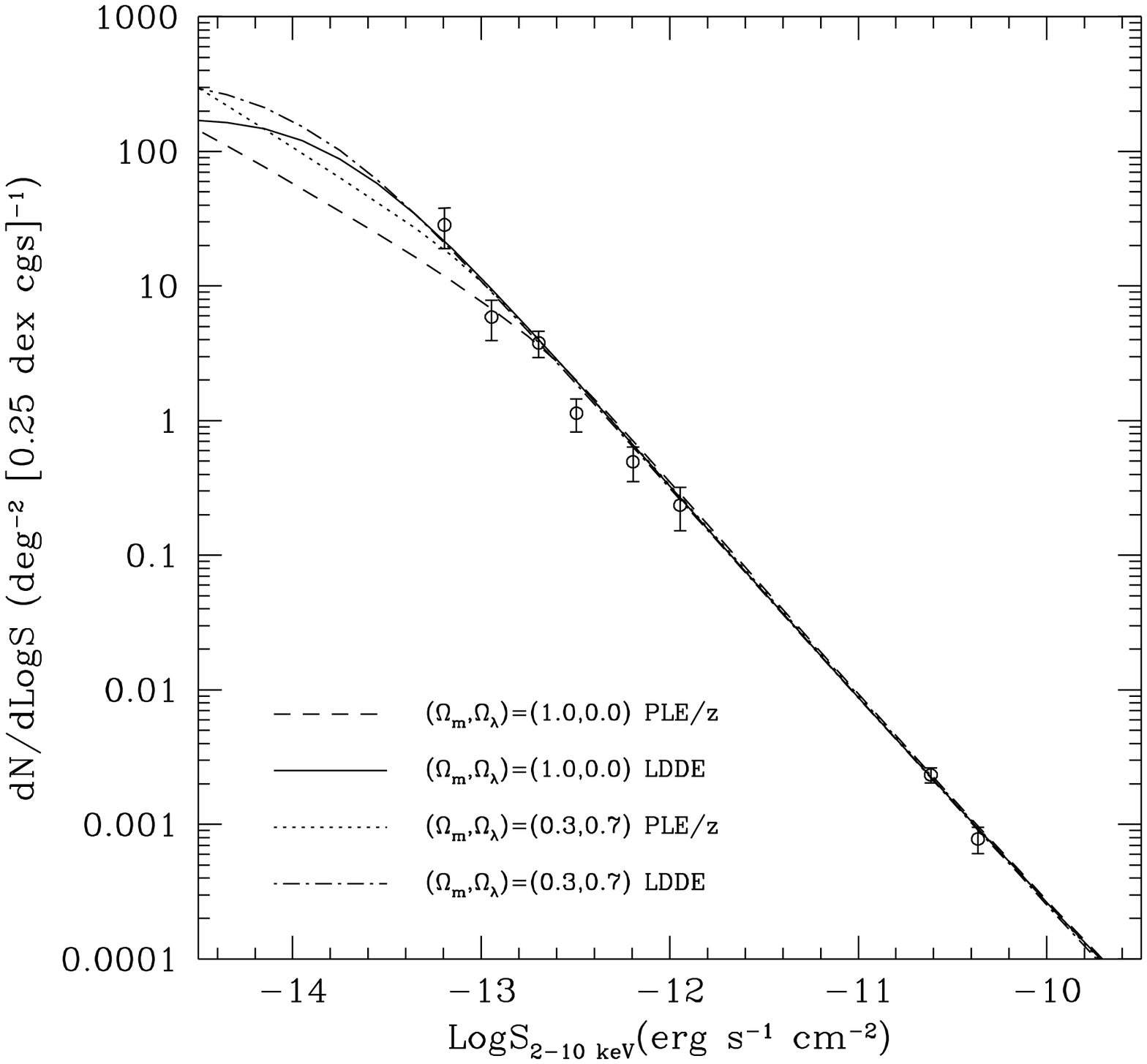,angle=0,width=8.1cm}
\caption{
The differential counts of the type 1 AGN used in the estimate of the
luminosity function. The models are explained in Table 6.
}
\addtolength{\baselineskip}{5pt}
\end{inlinefigure}

\section{Discussion and conclusions}

Thanks to the identification of 61 sources of the HELLAS sample we
have been able to double the number of hard X-ray AGN1 available for
statistical analysis at fluxes in the range $f_{2-10 keV}\sim
10^{-13.5}$--$10^{-12}$ \ecs. In total we can use 74 AGN1
at these fluxes (37 from HELLAS), which combined with the local sample
of Grossan have allowed to show directly the shape of the LF of AGN1
as function of redshift and measure its evolution.

The PLE models provide satisfactory fits of the data both in the
($\Omega_m$,$\Omega_\lambda$)=(1.0,0.0) and in the
($\Omega_m$,$\Omega_\lambda$)=(0.3,0.7) cosmologies. Our estimate of
the LF in the ($\Omega_m$,$\Omega_\lambda$)=(1.0,0.0) has a
significantly larger normalization in comparison to the previous
measure from Boyle et al. (1998).

The data start to probe in the hard X-rays the faint part of the LF
where the excess of density of AGN1 has been observed in the soft
X-rays, justifying the implementation of the LDDE models. However, in
both cosmologies, the statistic is not significant enough to
distinguish between the PLE and LDDE models (see Table 6).  In fact,
in the prediction of the differential counts of AGN1 shown in Figure
11, the models differentiates at fluxes fainter than
$f_{2-10keV}\sim10^{-13}$ \ecs, where the statistic is
still poor. The new upcoming fainter surveys from {\it Chandra} and
{\it XMM-Newton} will easily test which model is correct.

%
%
\begin{inlinefigure}
\figurenum{12}
\psfig{figure=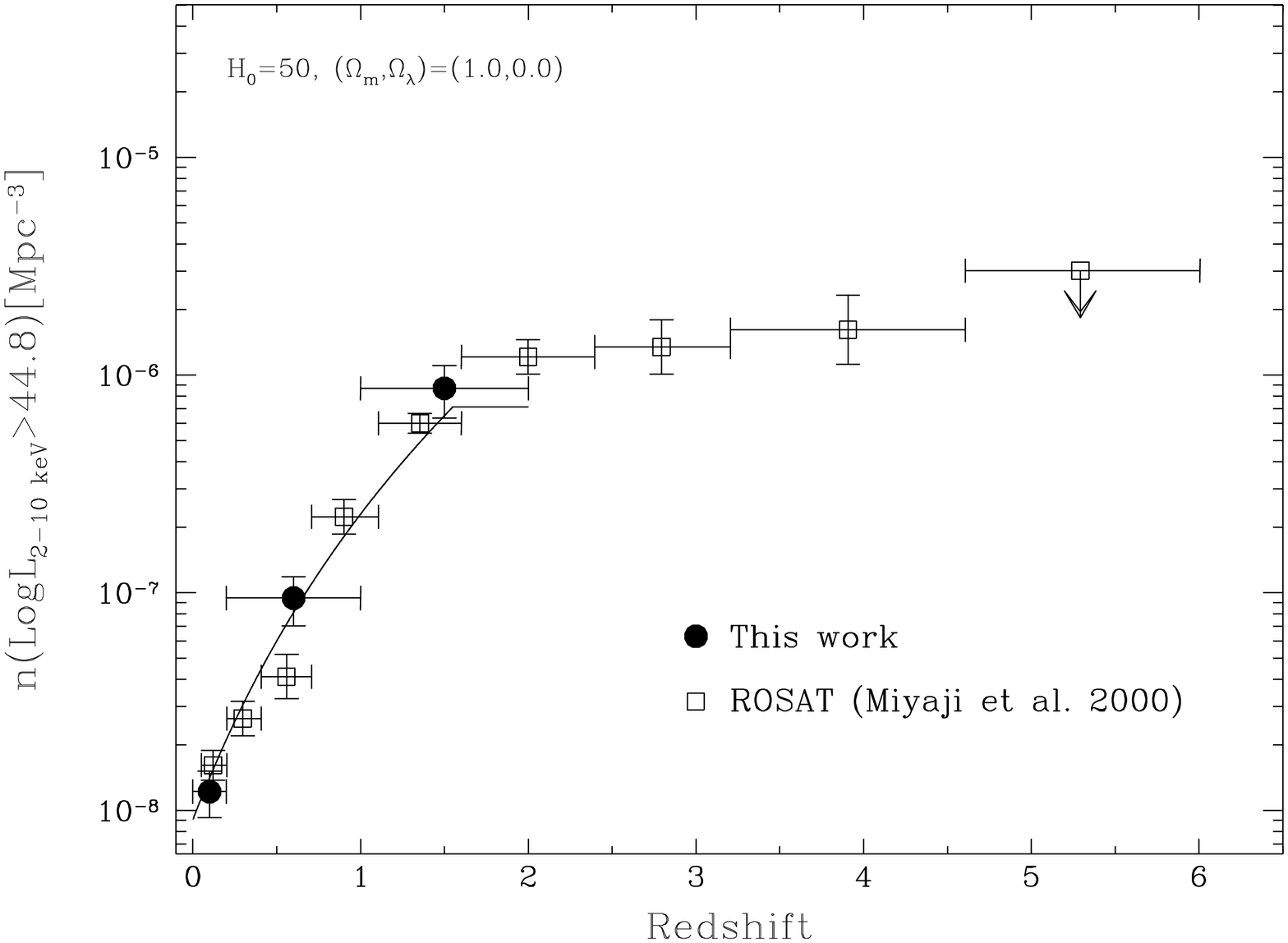,angle=-90,width=8.1cm}
\caption{
The evolution of the density of type 1 AGN having LogL$_X$(2-10
keV)$>$44.8 as a function of redshift. The soft X-ray data from
Miyaji et al. (2000) have been over-plotted assuming a slope
$\alpha=0.6$, which corresponds to a limit LogL$_X$(0.5-2 keV)$>$44.5.
The continuous line are the predictions of our LDDE model.
}
\addtolength{\baselineskip}{5pt}
\end{inlinefigure}

In Table 6 the percentages of the contribution to the 2-10 keV X-ray
background are shown. The X-ray background has been computed
integrating the LF up to $z=3.5$ for $L_X$$>$10$^{42}$ $erg/s$.  At
variance with the results of 24\% obtained from the evolution of the
LF derived by Boyle et al. (1998), our models reproduces from
$\sim$35\% up to $\sim$60\% of the XRB.  We used $I_{2-10}= 1.95\times
10^{-11}$ erg cm$^{-2}$ s$^{-1}$ deg$^{-2}$ from Chen, Fabian and
Gendrau (1997). The highest percentages would probably imply that part
of the absorbed population necessary to reproduce the XRB is already
included in the AGN1 at high redshift.  However, more detailed
analysis of this issue are beyond the scope of this paper. The
percentages would decreas if a value f $I_{2-10}=2.35
\times 10^{-11}$ erg cm$^{-2}$ s$^{-1}$ deg$^{-2}$ from Vecchi et
al. (1999) is assumed.

It is interesting to notice that AGN1 in the 2-10 keV range show an
evolution up to $z\sim2$ which is fairly well compatible with what
observed in the soft X-rays. In fact we found a good fit of the data
in the ($\Omega_m$,$\Omega_\lambda$)=(1.0,0.0) Universe if we assume
exactly the same parameters found in the soft X-rays by Miyaji et
al. (2000) for AGN1, by only looking for a fit with the break
luminosity $L_\ast$ (model 7 in Table 6). Miyaji et al. (2000) found
Log$L_\ast$=43.78. The value $L_\ast$ found by our fit in the 2-10 keV
band is Log$L_\ast$=44.13. This luminosity difference implies an X-ray
spectrum for AGN1 with slope $\alpha$=0.6, which is the same slope
used in our computation of the LF.  We note however that Miyaji et
al. (2000) used a slope $\alpha$=1.0 in their computations of the soft
X-ray LF. Therefore, taken at a face value, the match between the soft
and hard X-rays LFs implies that AGN1 have a broad band concave
spectrum getting steeper going toward lower energies.  As already
discussed in the previous section, the
($\Omega_m$,$\Omega_\lambda$)=(0.3,0.7) cosmology for AGN1 alone has
not been analyzed by Miyaji et al. (2000).

The agreement of the evolution measured in the soft and hard X-rays is
also shown in Figure 12, where the evolution of the density of AGN1
brighter than LogL$_{2-10 keV}$=44.8 is shown. The soft X-ray data
from Miyaji et al. (2000) have been over-plotted assuming a slope
$\alpha=0.6$, which corresponds to a limit LogL$_X$(0.5-2 keV)$>$44.5.
The continuous line are the predictions of our LDDE model
for ($\Omega_m$,$\Omega_\lambda$)=(1.0,0.0).

\acknowledgements
Based on observations collected at the European Southern Observatory,
Chile, ESO N$^{\circ}$: 62.P-0783, 63.O-0117(A), 64.O-0595(A), 65.O-0541(A).
This research has made use of the NASA/IPAC Extragalactic Database
(NED) which is operated by the Jet Propulsion Laboratory, California
Institute of Technology, under contract with the National Aeronautics
and Space Administration.  This research has been partially supported
by ASI contract ARS-99-75, MURST grants Cofin-98-02-32, Cofin-99-034,
Cofin-00-02-36 and a 1999 CNAA grant.

\appendix
\section{The biases introduced by the cross correlations with catalogues}

As already described, 25 sources have been identified by
cross-correlation with
existing catalogues. However, as the sample is not fully spectroscopic
identified, the cross-correlation with the existing catalogues could
alter the average characteristics (fluxes, percentages of classes of
counterparts, etc.) of the subsample of identified sources.
Namely, the subsample could not be representative of the whole sample.

Let's explain this with an example. In an ideal sample of 50 AGN1 and
50 AGN2 a random identification of a subsample of 40 sources will
identify about 50$\%$ AGN1 and 50$\%$ AGN2 according to sampling
errors.  But if a cross-correlation is first made with a catalogue of
only AGN1 (let say 30) and later only 10 sources are randomly
identified, the fraction of AGN1 in the total identified subsample of
40 sources will be artificially increased (they will be at minimum
75$\%$). Our subsample is risking the same sort of bias, and we wish
to quantify it.
  
A Kolmogorov-Smirnov (KS) two-sample test gives a 9$\%$ probability
that the X-ray flux distribution of the 25 sources identified through
the cross-correlation with catalogues belong to the same parent
population of the total sample of 118 sources.
Although not significant, this low probability is due to the average
slightly brighter fluxes of the cross-correlated subsample ($<{\rm
Log}F_{5-10~keV}>=-12.4\pm0.4$) in comparison to the total sample
($<{\rm Log}F_{5-10~keV}>=-12.6\pm0.3$). We have thus tried to
populate the faintest bins during the observing runs at the telescope,
and, indeed, the KS test gives a 66$\%$ probability that the whole
subsample of the 74 identified sources (included the empty fields)
belong to the same parent population of the total sample of 118
sources (see Figure 1).

In this way, our observing runs at the telescope have recovered the
possible alteration on the fraction of classes of sources which are
identified. As the catalogues used for the cross correlation are
mainly populated by AGN1 (we used NED), their fraction in our sample,
as explained before, could be artificially increased. AGN1 are 14 out
of the 25 cross-correlated sources (56$\%$). The fraction of AGN1 is,
as expected (but not significantly), lower for the sources observed at
the telescope: 23 out of the 49 (47$\%$). The two values are not
statistically distinguishable from the observed fraction 50$\%$ of
AGN1 in the whole sample, which we thus consider representative
of the whole HELLAS sample.

%
%
\begin{deluxetable}{cr}
\renewcommand\baselinestretch{1.0}
\tablewidth{0pt}
%
%
%
\parskip=0.2cm
\tablenum{1}
\scriptsize
\tablecaption{Sky coverage}
\tablehead{
Log(Flux) 5-10 keV & Area\cr
\ecs & deg$^2$
}
\startdata
 -10.34 & 55.51 \cr
 -11.85 & 55.51 \cr
 -11.89 & 55.48 \cr
 -11.93 & 55.39 \cr
 -11.97 & 55.28 \cr
 -12.00 & 55.05 \cr
 -12.04 & 54.66 \cr
 -12.08 & 54.04 \cr
 -12.11 & 53.16 \cr
 -12.15 & 51.97 \cr
 -12.19 & 50.63 \cr
 -12.22 & 49.04 \cr
 -12.26 & 47.13 \cr
 -12.30 & 44.88 \cr
 -12.34 & 42.44 \cr
 -12.37 & 39.93 \cr
 -12.41 & 37.49 \cr
 -12.45 & 34.88 \cr
 -12.48 & 31.93 \cr
 -12.52 & 28.83 \cr
 -12.56 & 25.67 \cr
 -12.59 & 22.75 \cr
 -12.63 & 20.03 \cr
 -12.67 & 17.53 \cr
 -12.70 & 15.23 \cr
 -12.74 & 13.16 \cr
 -12.78 & 11.32 \cr
 -12.82 &  9.73 \cr
 -12.85 &  8.30 \cr
 -12.89 &  6.96 \cr
 -12.93 &  5.67 \cr
 -12.96 &  4.56 \cr
 -13.00 &  3.60 \cr
 -13.04 &  2.88 \cr
 -13.07 &  2.35 \cr
 -13.11 &  1.96 \cr
 -13.15 &  1.60 \cr
 -13.19 &  1.27 \cr
 -13.22 &  0.94 \cr
 -13.26 &  0.68 \cr
 -13.30 &  0.47 \cr
 -13.33 &  0.33 \cr
 -13.37 &  0.22 \cr
 -13.41 &  0.14 \cr
 -13.44 &  0.08 \cr
 -13.48 &  0.05 \cr
 -13.52 &  0.03 \cr
 -13.56 &  0.01 \cr
 -13.59 &  0.00 \cr
\enddata
\label{tab1}
\end{deluxetable}

\bigskip

\begin{deluxetable}{llrrrrrrr}
\renewcommand\baselinestretch{1.0}
\tablewidth{0pt}
%
%
%
\parskip=0.2cm
\tablenum{5}
\footnotesize
\tablecaption{Emission Line Measurements}
\tablehead{
Name & $z~~~$ & Type & MgII &  OII & H$_{\beta}$ & OIII & H$_{\alpha}$+NII & SII \\
     &      &        & (\AA)&(\AA)&(\AA)&(\AA)&(\AA)&(\AA)
}
\startdata
H002636$-$194416             & 0.238 & FWZI   &     ... &   62.1 &   38.1 &   62.1 &      ... &      ... \\ 
                           &       & FWHM   &     ... &   19.7 &   15.4 &   18.5 &      ... &      ... \\ 
                           &       & EW     &     ... &   82.4 &   14.6 &  133.1 &      ... &      ... \\ 
\\
H004546$-$251550           & 0.111 & FWZI   &     ... &    ... &    ... &   40.1 &  104.2 &   44.1 \\ 
                         &       & FWHM   &     ... &    ... &    ... &   14.6 &   69.7 &   50.4 \\ 
                         &       & EW     &     ... &   -1.3 &   -0.2 &    5.6 &   22.2 &    2.5 \\ 
\\
H012157$-$584442           & 0.118 & FWZI   &     ... &   44.1 &   38.1 &   64.1 &  120.2 &   48.1 \\ 
                         &       & FWHM   &     ... &   19.7 &   13.1 &   16.8 &   45.0 &   80.0 \\ 
                         &       & EW     &     ... &   30.4 &    2.9 &   21.4 &   53.5 &   20.8 \\ 
\\
H013434$-$295816           & 2.217 &FWZI   &$^{a}$452.7 &      ... &      ... &      ... &      ... &      ... \\ 
                         &       & FWHM   &$^{a}$123.6 &      ... &      ... &      ... &      ... &      ... \\ 
                         &       & EW     &$^{a}$156.5 &      ... &      ... &      ... &      ... &      ... \\ 
\\
H013533$-$295202           & 1.344 & FWZI   & 282.5 &      ... &      ... &      ... &      ... &      ... \\ 
                         &       & FWHM   &  67.6 &      ... &      ... &      ... &      ... &      ... \\ 
                         &       & EW     &  61.2 &      ... &      ... &      ... &      ... &      ... \\ 
\\
H033408$-$360403           & 0.904 & FWZI   & 412.6 &   40.1 &      ... &      ... &      ... &      ... \\ 
                         &       & FWHM   & 174.0 &   26.5 &      ... &      ... &      ... &      ... \\ 
                         &       & EW     & 120.5 &    6.1 &      ... &      ... &      ... &      ... \\ 
\\
H043712$-$473148           & 0.142 &  FWZI   &     ... &   36.0 &  122.2 &   64.1 &  162.3 &   70.1 \\ 
                         &       &  FWHM   &    ... &   16.4 &   36.4 &   20.9 &   39.7 &   73.0 \\ 
                         &       & EW     &     ... &    5.2 &   65.4 &   23.6 &  241.2 &   17.0 \\ 
\\
H043847$-$472802           & 1.453 &  372.6 &      ... &      ... &      ... &      ... &      ... \\ 
                         &       &   64.9 &      ... &      ... &      ... &      ... &      ... \\ 
                         &       &   61.4 &      ... &      ... &      ... &      ... &      ... \\ 
\\
H064638$-$441534           & 0.153 & FWZI   &     ... &   60.1 &  128.2 &   60.1 &  122.2 &   46.1 \\ 
                         &       & FWHM   &     ... &   18.3 &   47.2 &   21.3 &   36.1 &   29.0 \\ 
                         &       &EW     &      ... &    7.9 &   61.0 &   78.9 &  168.4 &   10.0 \\ 
\\
H083737$+$254752           & 0.077 & FWZI   &     ... &   52.1 &   69.4 &   56.8 &  327.2 &   60.1 \\ 
                         &       & FWHM   &     ... &   40.3 &   28.8 &   15.2 &   71.8 &   43.0 \\ 
                         &       & EW     &     ... &   51.8 &   16.0 &   11.3 &  291.6 &   17.0 \\ 
\\
H083859$+$260814           & 0.048 & FWZI   &     ... &    ... &   36.7 &   40.1 &  110.2 &   70.1 \\ 
                         &       & FWHM   &     ... &    ... &   11.7 &   23.2 &   42.1 &   32.2 \\ 
                         &       & EW     &     ... &   -8.9 &    1.9 &    4.1 &   46.0 &    6.7 \\ 
\\
H103216$+$505120           & 0.174 & FWZI   &     ... &   30.0 &  203.7 &   36.7 &  150.2 &   33.4 \\ 
                         &       & FWHM   &     ... &   12.0 &   68.1 &   13.2 &   79.9 &   15.0 \\ 
                         &       & EW     &     ... &    5.8 &   65.1 &   13.1 &  163.2 &    6.5 \\ 
\\
H111814$+$402838           & 0.387 & FWZI   & 340.5 &   50.1 &    ... &   40.1 &      ... &      ... \\ 
                         &       & FWHM   & 184.0 &   46.7 &    ... &   10.3 &      ... &      ... \\ 
                         &       & EW     & 393.8 &    6.7 &   -0.7 &    6.7 &      ... &      ... \\ 
\\
H111849$+$402648           & 1.129 & FWZI   &$^{a}$200.3 &      ... &      ... &      ... &      ... &      ... \\ 
                         &       & FWHM   &$^{a}$65.6 &      ... &      ... &      ... &      ... &      ... \\ 
                         &       & EW     &$^{a}$59.9 &      ... &      ... &      ... &      ... &      ... \\ 
\\
H121853$+$295902           & 0.176 &FWZI   &      ... &   43.4 &   36.1 &   56.8 &  153.6 &   50.1 \\ 
                         &       &FWHM   &      ... &   12.4 &   12.1 &   12.5 &   26.8 &   20.9 \\ 
                         &       & EW     &      ... &   25.4 &   17.8 &  199.6 &  186.1 &   22.6 \\ 
\\
H124028$-$051402           & 0.300 & FWZI   &      ... &   80.1 &  166.9 &   63.4 &      ... &      ... \\ 
                         &       & FWHM   &      ... &   25.6 &   61.3 &   23.9 &      ... &      ... \\ 
                         &       & EW     &      ... &   36.1 &   90.8 &   67.1 &      ... &      ... \\ 
\enddata
\label{tab5}
\tablecomments{Observed frame.
Negative values correspond to absorption lines;
(a) C IV line.
}
\end{deluxetable}

\begin{deluxetable}{llrrrrrrr}
\renewcommand\baselinestretch{1.0}
\tablewidth{0pt}
\parskip=0.18cm
\tablenum{5}
\footnotesize
\tablecaption{Emission Line Measurements}
\tablehead{
Name & $z~~~$ & Type & MgII & OII & H$_{\beta}$ & OIII & H$_{\alpha}$+NII & SII \\
     &      &        & (\AA)&(\AA)&(\AA)&(\AA)&(\AA)&(\AA)
}
\startdata
H124036$-$050752           & 0.008 & FWZI   &      ... &  ... &   20.0  & -10.0 & 90.1 &     76.8 \\ 
                         &       & FWHM   &      ... &    ... &   13.9  &  -7.0 & 34.0 &     26.0 \\ 
                         &       & EW     &      ... &    ... &    0.8  &  -0.4 & 26.3 &      6.8 \\ 
\\
H130436$-$101549           & 2.386 & FWZI   &$^a$468.7 &      ... &      ... &      ... &      ... &      ... \\ 
                         &       & FWHM   &$^{a}$128.4 &      ... &      ... &      ... &      ... &      ... \\ 
                         &       & EW     &$^{a}$537.0 &      ... &      ... &      ... &      ... &      ... \\ 
\\
H134820$-$301156           & 0.128 & FWZI   &      ... &   52.1 &    ... &   56.1 &  112.2 &      ... \\ 
                         &       & FWHM   &      ... &   21.5 &    ... &   22.0 &   53.9 &      ... \\ 
                         &       & EW     &      ... &   19.0 &   -0.4 &   12.7 &   63.0 &      ... \\ 
\\
H134845$-$302946           & 0.330 & FWZI   &      ... &   36.0 &  220.4 &   72.1 &      ... &      ... \\ 
                         &       & FWHM   &      ... &   15.8 &   83.0 &   20.5 &      ... &      ... \\ 
                         &       & EW     &      ... &    2.3 &   84.2 &   55.8 &      ... &      ... \\ 
\\
H135015$-$302010           & 0.074 & FWZI   &      ... &   60.1 &    ... &   52.1 &  168.3 &   56.1 \\ 
                         &       & FWHM   &      ... &   23.1 &    ... &   24.5 &   64.4 &   31.7 \\ 
                         &       & EW     &      ... &  100.5 &   -0.6 &   16.0 &   48.8 &    8.0 \\ 
\\
H135354$+$182016           & 0.217 & FWZI   &      ... &   52.1 &   24.0 &   64.1 &  252.4 &   40.1 \\ 
                         &       & FWHM   &      ... &   13.7 &    9.3 &   19.7 &   95.0 &   47.9 \\ 
                         &       & EW     &      ... &   47.3 &    2.1 &   22.8 &  154.0 &   12.4 \\ 
\\
H151934$+$653558           & 0.044 & FWZI   &      ... &   36.0 &    ... &   44.1 &   92.2 &   44.1 \\ 
                         &       & FWHM   &      ... &   13.6 &    ... &    9.7 &    9.0 &   14.5 \\ 
                         &       & EW     &      ... &   43.6 &   -1.3 &   22.6 &   26.4 &    6.1 \\ 
\\
H163419$+$594504           & 0.341 & FWZI   &  144.2 &   48.1 &   72.1 &   56.1 &      ... &      ... \\ 
                         &       & FWHM   &   19.9 &   12.9 &   12.9 &   13.7 &      ... &      ... \\  
                         &       & EW     &   33.1 &   19.7 &   30.8 &  267.2 &      ... &      ... \\ 
\\
H165043$+$043618           & 0.031 & FWZI   &      ... &   48.1 &   36.1 &   52.1 &  124.2 &   56.1 \\ 
                         &       & FWHM   &      ... &   40.1 &   27.2 &   21.0 &   49.1 &   32.6 \\ 
                         &       & EW     &      ... &   45.9 &    2.8 &   26.6 &   43.0 &   12.3 \\ 
\\
H165238$+$022206           & 0.395 & FWZI   &  184.3 &   56.1 &  164.3 &   64.1 &      ... &      ... \\ 
                         &       & FWHM   &   70.0 &   24.1 &  129.5 & 21.5 &      ... &      ... \\ 
                         &       & EW     &   66.6 &   22.5 &   37.5 &   68.5 &      ... &      ... \\ 
\\
H204253$-$103826           & 0.363 & FWZI   &      ... &   72.1 &  184.3 &   76.1 &      ... &      ... \\ 
                         &       & FWHM   &      ... &   22.5 &   48.5 &   22.3 &      ... &      ... \\ 
                         &       & EW     &      ... &   15.4 &   73.0 &  201.8 &      ... &      ... \\ 
\\
H204435$-$102808           & 2.755 & FWZI   &$^{a}$248.4 &      ... &      ... &      ... &      ... &      ... \\ 
                         &       & FWHM   &$^{a}$101.9 &      ... &      ... &      ... &      ... &      ... \\ 
                         &       & EW     &$^{a}$56.6 &      ... &      ... &      ... &      ... &      ... \\ 
\\
H222632$+$211138           & 0.260 & FWZI   &      ... &      ... &  125.5 &   26.7 &      ... &      ... \\ 
                         &       & FWHM   &      ... &      ... &   41.6 &    7.9 &      ... &      ... \\ 
                         &       & EW     &      ... &      ... &   36.4 &    9.3 &      ... &      ... \\ 
\\
H231932$-$424228           & 0.101 & FWZI   &      ... &   68.1 &   40.1 &   56.1 &  172.3 &   64.1 \\ 
                         &       & FWHM   &      ... &   23.1 &   26.3 &   21.8 &   46.9 &   38.4 \\ 
                         &       & EW     &      ... &   20.4 &    5.0 &   24.2 &   77.6 &    7.5 \\ 
\\
H232729$+$084926           & 0.154 & FWZI   &      ... &   28.0 &   92.2 &   60.1 &  116.2 &   56.1 \\ 
                         &       & FWHM   &      ... &   15.9 &   80.3 &   18.7 &   34.2 &   37.5 \\ 
                         &       & EW     &      ... &   13.0 &   16.8 &   19.1 &   60.9 &   16.2 \\ 
\\
H232906$+$083416             & 0.953 & FWZI   &  368.6 &      ... &      ... &      ... &      ... &      ... \\ 
                           &       & FWHM   &   76.1 &      ... &      ... &      ... &      ... &      ... \\ 
                           &       & EW     &  133.6 &      ... &      ... &      ... &      ... &      ... \\ 
\\
H233154$+$193836             & 0.475 & FWZI   &  344.5 &      ... &      ... &      ... &      ... &      ... \\ 
                           &       & FWHM   &   75.0 &      ... &      ... &      ... &      ... &      ... \\ 
                           &       & EW     &   93.5 &      ... &      ... &      ... &      ... &      ... \\ 
\enddata
\label{tab5}
\tablecomments{Observed frame.
Negative values correspond to absorption lines;
(a) C IV line.
}
\end{deluxetable}
\newpage

\begin{figure*}
\figurenum{3}
\epsfxsize=160truemm
\epsffile{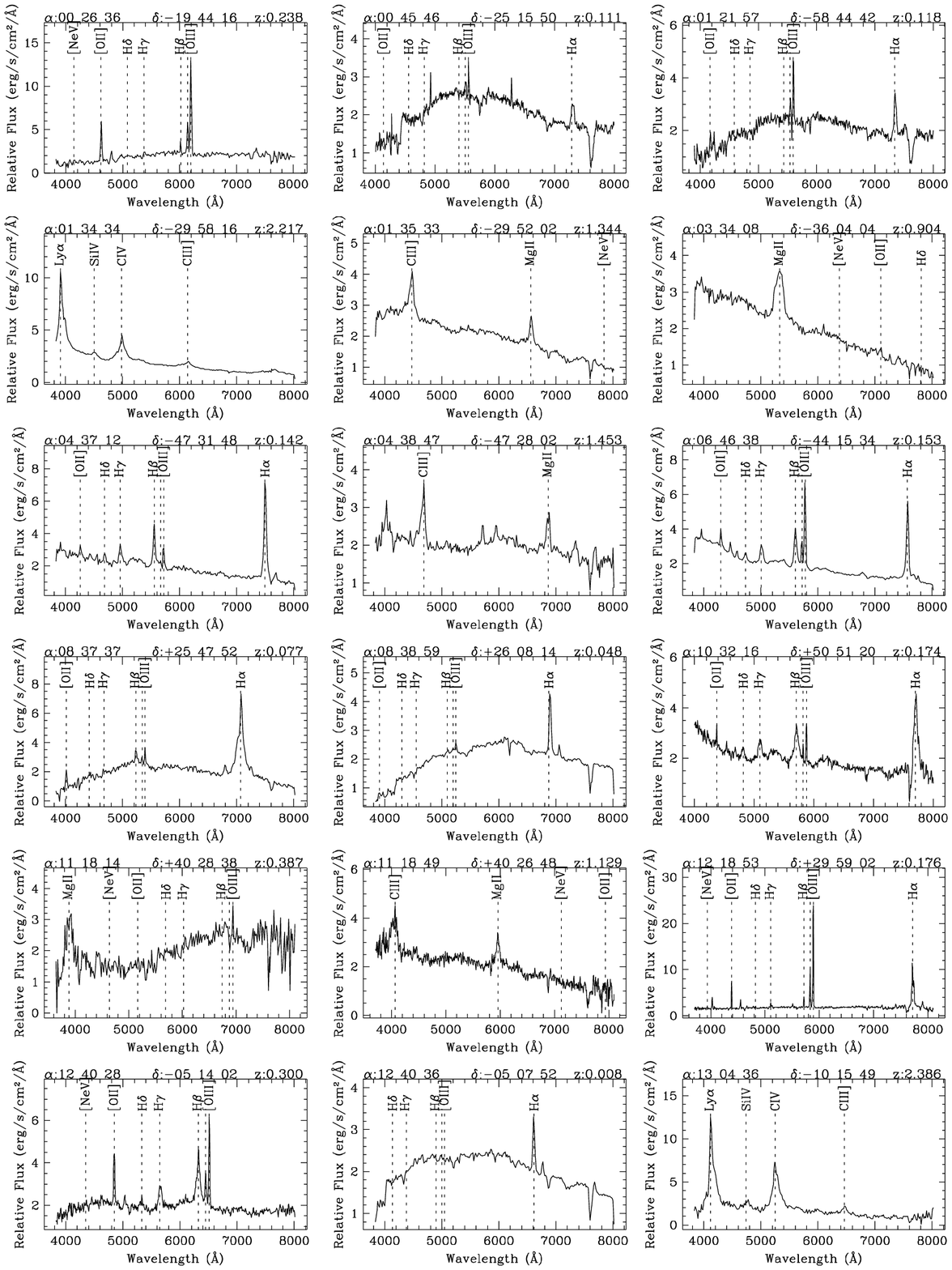}
\caption{The optical spectra of the most probable identified counterparts of the HELLAS sources.
The most typical emission lines for AGN are over-plotted with the corresponding
redshift.}
\label{f3}
\end{figure*}

\vfill\eject
\begin{figure*}
\figurenum{3}
\epsfxsize=160truemm
\epsffile{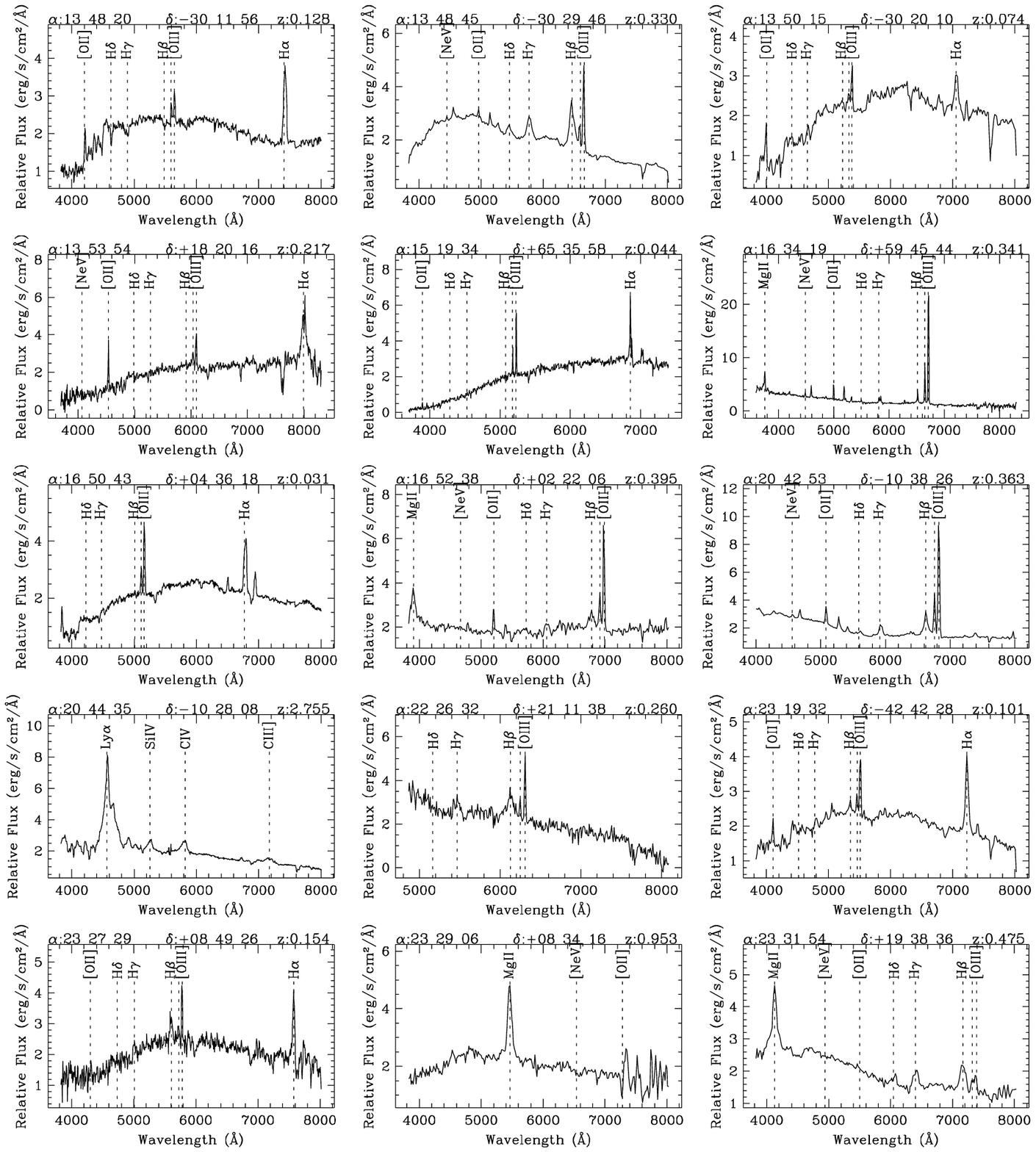}
\caption{continued}
\end{figure*}

\end{document}